\RequirePackage{fix-cm}
\documentclass[smallextended]{svjour3}       %
\smartqed 

\usepackage{amsmath,amssymb}
\usepackage{bm}
\usepackage{graphicx}
\usepackage{dcolumn}
\usepackage{qcircuit}
\usepackage{epstopdf}
\usepackage[caption=false]{subfig}
\usepackage[numbers,sort&compress]{natbib}
\usepackage{color}
\usepackage{float}
\usepackage[colorlinks=true,linkcolor=blue,citecolor=red, linktocpage=true,breaklinks=true]{hyperref}

\newcommand{\tr}{\mathrm{Tr}}

\newcolumntype{Y}{>{\centering\arraybackslash}X}
\begin{document}

\title{Implementation of efficient quantum search algorithms on NISQ computers}

\author{Kun Zhang \and
        Pooja Rao \and
        Kwangmin Yu \and
        Hyunkyung Lim \and
        Vladimir Korepin \and
}

\institute{Corresponding Author: K. Zhang \at
              Department of Chemistry, State University of New York at Stony Brook, Stony Brook, New York 11794, USA  \\
              \email{kun.h.zhang@stonybrook.edu}
           \and
           P. Rao \at
           Mathematical Sciences Research Institute, Berkeley, California 94720, USA \\
           \email{prao@msri.org}
           \and
           K. Yu \at
           Computational Science Initiative, Brookhaven National Laboratory, Upton, New York 11973, USA \\
           \email{kyu@bnl.gov}
           \and
           H. Lim \at
           Department of Applied Mathematics and Statistics, Stony Brook University, Stony Brook, New York 11794, USA \\
           \email{hyun-kyung.lim@stonybrook.edu}
           \and
           V. Korepin \at
           C.N. Yang Institute for Theoretical Physics, Stony Brook university, Stony Brook, New York 11794-3840, USA\\
           \email{vladimir.korepin@stonybrook.edu}
}

\date{Received: date / Accepted: date}

\maketitle

\begin{abstract}
	
	Despite the advent of Grover's algorithm for the unstructured search, its successful implementation on near-term quantum devices is still limited. We apply three strategies to reduce the errors associated with implementing quantum search algorithms. Our improved search algorithms have been implemented on the IBM quantum processors. Using them, we demonstrate three- and four-qubit search algorithm with higher average success probabilities compared to previous works. We present the successful execution of the five-qubit search on the IBM quantum processor for the first time. The results have been benchmarked using degraded ratio, which is the ratio between the experimental and the theoretical success probabilities. The fast decay of the degraded ratio supports our divide-and-conquer strategy. Our proposed strategies are also useful for implementation of quantum search algorithms in the post-NISQ era.
	
	
\end{abstract}

\section{\label{sec:intro}Introduction}

In recent years, much progress has been made in building quantum processors \cite{Barends14,BHLSL16,FMLLDM17,Google20} and demonstrating quantum advantage \cite{Arute19,Zhong20}. In fact, quantum algorithms are the reason why quantum computers are so powerful \cite{NC10}. However, the errors resulting from noisy quantum gates and decoherence make these devices far from perfect. The term ``Noisy Intermediate-Scale Quantum'' (NISQ) has been coined to describe the current era of noisy quantum computers \cite{Preskill18}.


Circuit depth is a practical metric for quantum circuits. Circuit depth is defined as the number of consecutive elementary operations required to run a circuit on quantum hardware. For the same circuit, different hardwares may give different depths since the connectivity may vary from machine to machine \cite{Cross19}. Most quantum computers have the elementary single- and two-qubit gates, but the running time of an algorithm on a quantum computer is directly related to the number of two-qubit gates. The two-qubit gates are much harder to realize in experiments (also take more time than single-qubit gates) since they create entanglement and the states become classically intractable \cite{NC10}. Circuits with longer depths are more susceptible to gate and decoherence errors. Thus, NISQ era algorithms strive for shallow depths \cite{Bharti21}. 

Grover's algorithm is well-known for providing quadratic speedup for unstructured search problem \cite{Grover97,GK17}. It has wide applications, from exhaustive search for NP-hard problems \cite{BBBV96} to quantum machine learning \cite{Biamonte17}. The theoretical complexity of Grover's algorithm is based on number of queries to oracle, often referred to as the black box. The oracle can identify the target item in a database. Grover's algorithm has been proven to be strictly optimal in the number of queries to the oracle \cite{BBHT98,Zalka99}.

Theoretical computational cost measures based on the oracular complexity, although useful for the theoretical analyses, are not very practical for assessing the performance of a quantum algorithm on real quantum machines. Given the wide-range of applications of Grover's algorithm, a line of research has been directed towards estimating its implementation cost, including its depth and width requirements \cite{Grassl16,Kim18,Jaques20,Wang20}. While querying the oracle is an important operation, it is not the sole operation in Grover's algorithm. The other important part of Grover's algorithm is the diffusion operator. Previous studies have shown that variants of Grover's algorithm allow different choices for the diffusion operator, while maintaining the quantum speedup \cite{Kato05,Tulsi15,JRW17}. Partial diffusion operators, also called ``local" diffusion operators, act on a subspace of the database. They are the key components in the partial search algorithms \cite{GR05,Korepin05,KG06}. Interestingly, they can also be applied to the full search problem, decreasing the depth of the quantum search algorithms \cite{Grover02,Marcin20,Zhang20,Liu21}. Such realizations make them much more viable for the NISQ devices.

Grover's algorithm for up to four-qubit search domain ($2^4 = 16$ elements) has been implemented previously on the IBM quantum processors \cite{Mandviwalla18,Gwinner20,Satoh20} for unstructured search. In this paper, we apply three different strategies to improve the performance of quantum search algorithms on the NISQ devices: (i) the hybrid classical-quantum search, (ii) use of partial diffusion operators to optimize the depth of quantum search algorithms, and (iii) the divide-and-conquer search. Here the divide-and conquer means that we find the partial target string at each step. Since we are considering unstructured search problem, we do not recursively apply the divide-and-conquer strategy here. The three strategies can be jointly applied for an enhanced error mitigation. We demonstrate the improved three- and four-qubit search implementations over the standard Grover's algorithm. The success probabilities are higher than previous reported results \cite{Mandviwalla18,Gwinner20}. For the five-qubit cases on IBM quantum processors, the search results from the direct execution of Grover's algorithms are too noisy, no better than the random guess. Our improved version, based on the proposed hybrid classical-quantum strategy, gives higher success probabilities than the purely classical approach (classical linear search). To the best of our knowledge, this is the first time that the five-qubit search algorithm has been successfully executed on IBM quantum processors. Note that the five-qubit search has recently been implemented on the trapped-ion qubits \cite{Hlembotskyi20}. We benchmark our results using the degraded ratio of success probabilities. The fast decay of the degraded ratio observed in our results implies favoring of shallow depth circuits on IBM quantum processors.

This paper is organized as follows. In Sec.~\ref{sec:QSA}, we review full and partial search algorithms as well as introduce the notations used in our paper. In Sec.~\ref{sec:noise reduction}, we talk about three different strategies as mentioned above to improve the quantum search algorithms on real quantum devices. We present the results from executing these computational strategies on IBM's quantum computers in Sec.~\ref{sec:IBMQ experiment}. Lastly, the conclusions from our study are presented in Sec.~\ref{sec:conclusion}. Appendix includes more details on our notations and results presented in the main text.

\section{\label{sec:QSA}Quantum search algorithms}

First, we give a brief review of Grover's algorithm. Then we introduce the partial diffusion operator, a key component in our paper, that acts only on a subspace of the search domain. 

\subsection{\label{subsec:Grover}Grover's algorithm}

Grover's algorithm is realized by repeatedly applying the Grover operator, denoted as $G_n$, on the initial state $|s_n\rangle$ \cite{Grover97,GK17}. The symbol $n$ denotes the number of qubits, which implies that the number of items in the database is $N=2^n$. The initial state, $|s_n\rangle$, is uniform superposition of computational basis states of the Hilbert space, $\mathcal H_{2}^{\otimes n}$. It can be realized by applying the Hadamard gate, $H$, as \cite{NC10}
\begin{equation}
\label{def s n}
    |s_n\rangle=H^{\otimes n}|0\rangle^{\otimes n}.
\end{equation}
Note that such a highly nontrivial initial state can easily be prepared with depth one. 

The Grover operator, $G_n$, is a composition of two operators, the oracle and the diffusion operator. The oracle marks the target item and the diffusion operator creates an inversion about the mean. In Grover's algorithm, a query to the phase oracle results in a sign flip on the target state. We denote the oracle operation as
\begin{equation}
\label{def U t}
O_t=1\!\!1_{2^n}-2|t\rangle\langle t|. 
\end{equation}
Here, $|t\rangle$ is the target state representing the target string $t$. The target state $|t\rangle$ is also one of the computational basis states. We also refer to $t$ as the target item or the target string in our paper. Operator $1\!\!1_{2^n}$ is the identity operator acting on $\mathcal H_2^{\otimes n}$. For convenience, we assume that there is a unique target state in the database. However, the depth reduction strategies in the next section do not limit to the case with a unique target state. The diffusion operator is independent of the oracle, and is defined as  
\begin{equation}
\label{def I n}
D_n = 2|s_n\rangle\langle s_n|-1\!\!1_{2^n}.
\end{equation}
The oracle operator, $O_t$, can be viewed as a reflection in the plane perpendicular to the target state $|t\rangle$. The diffusion operator, $D_n$, reflects the amplitude in the average, since the state, $|s_n\rangle$, is the equal superposition of all items in the database. 

Composed of the oracle, $O_t$, and the diffusion operator, $D_n$, the Grover operator, $G_n$, is given by
\begin{equation}
\label{def G n}
G_n =  D_n O_t.
\end{equation}
Starting with the initial state, $|s_n\rangle$, and iteratively applying the Grover operator, $G_n$, on subsequent states, gives
\begin{equation}
\label{def P n} 
P_n(j) = |\langle t|G_n^j|s_n\rangle|^2 = \sin^2((2j+1)\theta),  
\end{equation}
where $P_n(j)$ is the probability finding the target string $t$ after $j$ iterations of the Grover operator on the initial state. The angle $\theta$ is defined as $\sin\theta=1/\sqrt N$. When $j$ reaches $j_\text{max}=\lfloor \pi\sqrt N/4\rfloor$, the probability approaches unity. Thus, the oracular complexity of Grover's algorithm is $\mathcal O(\sqrt N)$, which is quadratic speedup compared to the classical complexity, $\mathcal O(N)$. The idea behind Grover's algorithm is to increase the amplitude of the target state (approximately) linearly, which leads to a quadratic change in the probability as it is the amplitude squared. Moreover, Grover's algorithm is not limited to a specific initial state, such as the uniformly superimposed state, $|s_n\rangle$. As long as some distributions of the database can be efficiently realized, the amplitude of the target state can be amplified via Grover's algorithm. This general version of the algorithm is called the amplitude amplification algorithm \cite{Grover98,BHMT00}.

Although the general formalism of Grover's algorithm is simple, realizing it on real quantum computers (for unstructured search problems) is a non-trivial question. Different search problems have different realizations of the oracle. Recent studies have shown how to construct the oracle (via the elementary quantum gates) for the AES key search \cite{Grassl16,Almazrooie18,Langenberg19,Jaques20} and the MAX-CUT problem \cite{Satoh20}. The construction of diffusion operator on real quantum devices is more straightforward. The diffusion operator, $D_n$, and the $n$-qubit Toffoli gate denoted as $\Lambda_{n-1}(X)$, are single-qubit-gate equivalent \cite{NC10}. The notation $X$ denotes the NOT gate, while $n-1$ means that there are $n-1$ control qubits. For example, when $n=2$, $\Lambda_{n-1}(X)$ gives the CNOT gate. The $n$-qubit Toffoli gate $\Lambda_{n-1}(X)$ can be decomposed as a combination of single- and two-qubit gates with the depth linear in $n$ (with ancillary qubits) \cite{BBCDMSSSW95}.

\subsection{\label{subsec:PDO} Partial diffusion operator} 

The diffusion operator, $D_n$, defined in Eq.~(\ref{def I n}), reflects the amplitudes in the average of all items. We can generalize such an operator as
\begin{equation}
    \label{def I n m}
    D_{n,m} = 1\!\!1_{2^{n-m}}\otimes (2|s_m\rangle\langle s_m|-1\!\!1_{2^m}),
\end{equation}
with $m\leq n$. The diffusion operator, $D_{n,m}$, only reflects the amplitude in the subspace of the database. As $m<n$, we refer to $D_{n,m}$ as the local or partial diffusion operator. For convenience, we  drop $n$ from the notation to denote $D_{m} \equiv D_{n,m}$, without any possibility of confusion. Combined with the oracle operator, $O_t$, the local Grover operator is defined as
\begin{equation}
    \label{def G m}
    G_{m}=D_{m}O_t.
\end{equation}
Note that $G_{m}$ is still an $n$-qubit operator since the oracle acts on the full $n$-qubit space.

The local diffusion operator can naturally solve the partial search problem \cite{GR05,Korepin05,KG06}, which finds the substring of the target state. For example, the target state $|t\rangle$, can be decomposed as $|t\rangle=|t_1\rangle\otimes |t_2\rangle$. Assume that $t_1$ is $(n-m)$-bit length while $t_2$ is $m$-bit length. We can think that the database is divided into $K=2^{n-m}$ blocks. Each block has $b = 2^m$ number of items ($N=bK$). Also, each block has the partial target string $t_2$. The quantum partial search algorithm (QPSA) finds the target block represented by the target string $t_1$. The target string $t_2$ is not concerned.

The most efficient QPSA (based on the oracular complexity) starts by running the global Grover operators first, followed by running the local Grover operators and lastly, runs a single global Grover operator \cite{KL06}. Note that the operators, $G_n$ and $G_m$, do not commute \cite{KV06}. Thus, different orders of operators give different success probabilities. The QPSA trades accuracy for speed (based on the oracular complexity). The QPSA finds the target substring $t_1$, with fewer queries to the oracle than the full search algorithm. The reduced number of oracles, compared to Grover's algorithm, scales as $\sqrt{b}$ \cite{GR05,Korepin05,KG06}.

Although QPSA is the main application of the local diffusion operator, Grover introduced the local diffusion operator before the invention of QPSA. The local diffusion operator, introduced in \cite{Grover02}, aims to reduce the total number of gates in the quantum search algorithm. This motivation is easy to see as the partial diffusion operator, $D_m$, can be realized with fewer elementary quantum gates than the global diffusion operator, $D_n$, with $n>m$. Recent studies have revealed several other ways to reduce the depth of quantum search algorithm by exploiting the partial diffusion operator \cite{Marcin20,Zhang20}. We will discuss different strategies to improve the performance of quantum search algorithms on real devices in the next section.

\section{\label{sec:noise reduction}Strategies to improve quantum search algorithms}

In this section, we present three different strategies to improve accuracy and efficiency of the quantum search algorithm on the NISQ processors. Every strategy utilizes the local diffusion operator, $D_m$, as defined in Eq.~(\ref{def I n m}). In NISQ era, such strategies are important because they reduce the depth of the circuits. In post-NISQ era, such strategies can potentially reduce the physical resources needed for error correction, as well as the running time of the algorithms.

\subsection{\label{subsec:hybrid_search}Hybrid search algorithm}

The local diffusion operator, $D_m$, only acts on the subset of the given database. We can renormalize the search space in order to exploit $D_m$. Suppose that the search problem is to find the target string $t$ with length $n$. The oracle can only recognise the target state $|t\rangle$. By renormalizing the search space, we prepare the initial state $|t_1'\rangle\otimes|s_m\rangle$, where $t'_1$ is a specific string with length $(n-m)$. If $t_1'=t_1$ ($|t\rangle = |t_1\rangle\otimes|t_2\rangle$), then using Eq.~(\ref{def P n}), the probability of finding $t_2$ after $j$ iterations of $G_m$ on $|t_1\rangle\otimes|s_m\rangle$ is
\begin{equation}
    P_m(j) = \sin^2((2j+1)\theta_b),
\end{equation}
with $\sin\theta_b = 1/\sqrt b$ and $b=2^m$. 

Let the probability that $t_1'$ is $t_1$ be $P(t_1'=t_1)$. The probability $P_m(j)$ is conditioned on the probability $P(t_1'=t_1)$. Then the total probability of finding the target string $t$ with $j$ iterations of $G_m$ is
\begin{equation}
    P'_n(j) = P(t_1'=t_1)P_m(j).
\end{equation}
For the unstructured search problem, the classical probability, $P(t_1'=t_1)$, can only be given by randomly guessing on $2^{n-m}$ bits, i.e., $P(t_1'=t_1) = 1/2^{n-m}$. Random guessing does not require any quantum computational resources. We call such a search method as the hybrid classical-quantum search algorithm.

The advantage for the hybrid classical-quantum search algorithm is twofold. First, the depth of $G_m$ is smaller than $G_n$. If the total depth is fixed, the hybrid classical-quantum algorithm can apply more iterations of oracle. Second, quantum coherence during the algorithm is only required on the subspace $\mathcal H_2^{\otimes m}$. The full search algorithm is based on the coherence on $\mathcal H_2^{\otimes n}$, which is more fragile. Although the theoretical success probability is always smaller than the full search success probability with the same number of oracles, the real success probability could be higher because of its shorter depth and limited coherence between qubits. The theoretical success probability of hybrid search decays exponentially with respect to the number of randomly guessed qubits. Therefore, we do not expect a large number of randomly guessed qubits, especially for a large database. Therefore the hybrid strategy may not be suitable for post-NISQ search problems.


\subsection{\label{subsec:depth_opt}Depth optimization by partial diffusion operators}

Grover's algorithm is optimal in the number of queries to the oracle \cite{BBHT98,Zalka99}. However, the oracular complexity is not the only metric for determining an algorithm's requirement of the physical computational resources (such as the depth and the width of the circuit). The Grover operator is a combination of oracle operator and diffusion operator. The depth of the quantum search circuit can be reduced if we replace the global diffusion operator $D_n$ by the local diffusion operator $D_m$. There are different ways to do such replacements \cite{Grover02,Marcin20,Zhang20,Liu21}. Here, we follow the ideas from \cite{Zhang20}, which provides a general framework for depth optimization. 

Suppose that we design the search circuits by the operator
\begin{equation}
    S_{n,m}(\tilde j) = G_n^{j_1}G_m^{j_{2}}\cdots G_n^{j_{q-1}}G_m^{j_q},
\end{equation}
with $\tilde j=\{j_1,j_2,\ldots,j_q\}$. Every local diffusion operator acts on the same subspace. To remove the ambiguity of the notation $S_{n,m}(\tilde j)$, we require that the last number $j_q$ is always for the local Grover operator. For example, $S_{6,4}(\{2,0\}) = G_6^2$ and $S_{6,4}(\{1,1\}) = G_6G_4$. Note that $S_{n,m}(\{j,0\}) = G_n^j$ is the standard Grover's algorithm. We only consider one kind of local diffusion operator here. The idea below can be generalized to multi-type local diffusion operators local diffusion operators acting on different sizes of blocks, which gives the search operator $S_{n,m_1,m_2,\cdots,m_k}$.

The success probability of finding the target state by the operator $S_{n,m}(\tilde j)$ is
\begin{equation}
    P_{n,m}(\tilde j) = |\langle t|S_{n,m}(\tilde j)|s_n\rangle|^2.
\end{equation}
Since Grover's algorithm is strictly optimal in number of queries to the oracle \cite{Zalka99}, a local diffusion operator can only decrease the success probability compared to Grover's algorithm (with the fixed number of oracles). For example, $P_{n,m}(\{j_1,j_2\})\leq P_{n,m}(\{j_1+j_2,0\})$. 

The physical resources of quantum computers are the depth and the width. The depth roughly represents the physical running time of the circuit. We denote the depth of operator $U$ as $d(U)$. For the same operator, different devices may have different depths due to the different connectivity of the qubits and the different sets of universal gates. Operator $S_{n,m}(\tilde j)$ can have lower depth compared to $G_n^j$ (with the same number of oracles). For example, $d(S_{n,m}(\{j_1,j_2\}))\leq d(S_{n,m}(\{j_1+j_2,0\}))$. We introduce the expected depth of the search circuit $S_{n,m}(\tilde j))$ as
\begin{equation}
\label{def:d_exp}
\langle d_{n,m}(\tilde j)\rangle = \frac{d(S_{n,m}(\tilde j))}{P_{n,m}(\tilde j)}.
\end{equation}
Then the depth optimization strategy is to find the minimum of $\langle d_{n,m}(\tilde j)\rangle$ given by
\begin{equation}
\label{def d n}
    \langle d_{n}\rangle = \min_{m,\tilde j}\langle d_{n,m}(\tilde j)\rangle.
\end{equation}
We also optimize the size of the local diffusion operator given by the parameter $m$. Although we apply the local diffusion operator in $S_{n,m}(\tilde j)$, our algorithm is not the partial search algorithm.

The minimal expected number of oracles for Grover's algorithm is studied in \cite{BBHT98,GWC00}. Recall that the maximal iteration (giving the maximal success probability) is $j_\text{max}=\lfloor \pi\sqrt N/4\rfloor$. However, the minimal expected number of oracles is given by $j_\text{exp}=\lfloor 0.583\sqrt N \rfloor$, which is smaller than $j_\text{max}$. Incorporating the depth of global Grover operator $d(G_n)$, the optimal iteration number, $j_\text{exp}$, can give the minimal expected depth of Grover's algorithm. The significance of $\langle d_{n}\rangle$ (given by the partial diffusion operator) is to win over the minimal expected depth of Grover's algorithm. Theoretical study shows that there is a critical depth ratio (the ratio between the depths of the oracle and the global diffusion operator), below which Grover's algorithm is not optimal in depth \cite{Zhang20}. Such a critical depth ratio scales as $\mathcal O(n^{-1}2^{n/2})$. For example in the 10-qubit search, the second strategy can be applied if the depth of oracle is smaller than 83.97 times the depth of the 10-qubit Toffoli gate. In practice, we would not have such a large oracle depth. For example, based on the data in \cite{Jaques20}, the depth of AES-128 oracle is around 10 times the depth of global diffusion operator.

\subsection{\label{subsec:divde_conquery}Divide-and-conquer strategy}

NISQ devices can only run shallow depth circuits \cite{Preskill18}. Recent benchmarking results suggest that the fidelity of a circuit does not linearly decrease with the depth of two-qubit gates \cite{Gwinner20}. If we can reinitialize the input during the algorithm, then we can prevent the accumulation of errors at subsequent stages. The divide-and-conquer search algorithm is naturally related to QPSA. We can find partial bits of the target item, then renormalize the database to find the rest of the target string. 

For simplicity, we consider the two-stage quantum search algorithm. It can be easily generalized into the multi-stage search algorithm. In the first stage, the task is to find the target substring, $t_1$, with high probability. The second stage finds the rest  target string, $t_2$. The second-stage circuit will be dependent on the results from the first stage. The underlying idea behind the two-stage search algorithm is similar to the idea behind the hybrid classical-quantum search algorithm in Sec. \ref{subsec:hybrid_search}. The difference is that both the stages are realized by quantum search algorithms. Suppose that the first stage is realized by the operator $S^{(1)}_{n,m}(\tilde j)$. The initial state for the first stage is $|s_n\rangle$. The probability finding the target substring, $t_1$, is given by
\begin{equation}
    P^{(1)}_{n,m}(\tilde j) = \tr \left[\left(|t_1\rangle\langle t_1|\otimes 1\!\!1_{2^{m}}\right) S^{(1)}_{n,m}(\tilde j)|s_n\rangle\langle s_n|S^{(1)}_{n,m}(\tilde j)^\dag\right].
\end{equation}
The diffusion operator, $D_m$, in $S^{(1)}_{n,m}(\tilde j)$ acts on the qubits within the target substring, $t_2$. In other words, we measure the qubits which are not acted upon by $D_m$. Such a circuit design comes from QPSA. 

Suppose we find $t_1$ at the first stage. We prepare the initial state $|t_1\rangle\otimes |s_m\rangle$. Then design the operator $S^{(2)}_{n,m'}$ with $m'<m$ for the second stage. Such a circuit is the normalized version of the full search algorithm. The probability of finding the remaining target string $t_2$ is  
\begin{equation}
    P^{(2)}_{n,m}(\tilde j') = \tr \left[\left(1\!\!1_{2^{n-m}}\otimes|t_2\rangle\langle t_2|\right) S^{(2)}_{n,m'}(\tilde j')|t_1,s_m\rangle\langle t_1,s_m|S^{(2)}_{n,m'}(\tilde j')^\dag\right],
\end{equation}
with short notation $|t_1,s_m\rangle = |t_1\rangle\otimes |s_m\rangle$. The operator $S^{(2)}_{n,m'}$ does not change the initial state $|t_1\rangle$.

The expected depth of the above two-stage search algorithm is
\begin{equation}
\label{def:d_exp2}
\langle d_{n,m,m'}(\tilde j,\tilde j')\rangle = \frac{d(S^{(1)}_{n,m}(\tilde j))+d(S^{(2)}_{n,m'}(\tilde j'))}{P^{(1)}_{n,m}(\tilde j)P^{(2)}_{n,m}(\tilde j')}.
\end{equation}
The minimal expected depth can be obtained by optimizing the operators and the size of the diffusion operators:
\begin{equation}
    \langle d_{n,2}\rangle = \min_{m,m',\tilde j,\tilde j'}\langle d_{n,m,m'}(\tilde j,\tilde j')\rangle
\end{equation}
We add a subscript $2$ in $\langle d_{n,2}\rangle$ to distinguish the minimal expected depth of the full search $\langle d_{n}\rangle$ in Eq.~(\ref{def d n}). It is expected that $\langle d_{n,2}\rangle<\langle d_{n}\rangle$, since the measurement in the middle wipes out the amplified amplitude of the state $|t_2\rangle$. Only when the depth of the oracle is comparable to the depth of the global diffusion operator, the two-stage search algorithm can have lower depth than Grover's algorithm \cite{Zhang20}. 

The motivation for the multi-stage circuits is to mitigate the errors. Another advantage of the multi-stage search circuit is its ability to run the quantum search algorithm in parallel \cite{GWC00,Zhang20}. We can assign the first stage circuit to different quantum computers. Then each device finds a different part of the target string $t$. Combining all the results gives the full target string.

\section{\label{sec:IBMQ experiment}Implementation on IBM quantum processors}

First, we briefly present the basic setup as well as the circuit design from our implementation of the algorithms on the IBM quantum processors. Then we discuss the results on the three-, four-, and five-qubit search in the following subsections. 

\subsection{\label{subsec:circuit}Circuit designs}

The target item $t$ is encoded in the oracle. We have assumed the uniqueness of the target item. As toy model, we choose the phase oracle presented in \cite{FMLLDM17}. The $n$-qubit phase oracle is single-qubit-gate-equivalent to the $n$-qubit Toffoli gate $\Lambda_{n-1}(X)$ (or the $n$-qubit controlled phase gate $\Lambda_{n-1}(Z)$). Note that the diffusion operator $D_n$ is also single-qubit-gate-equivalent to the $n$-qubit Toffoli gate $\Lambda_{n-1}(X)$ \cite{NC10}. Although Qiskit provides the built-in $n$-qubit controlled gate, its fidelity and efficiency are not optimal. In the following, we show the realizations of three-, four-, and five-qubit controlled phase gate $\Lambda_{n-1}(Z)$ from our implementation. 

It is well-known that the three-qubit controlled phase gate, $\Lambda_{2}(Z)$ (or the Toffoli gate $\Lambda_{2}(X)$), can be realized by six CNOT gates, with full connectivity between the three qubits \cite{NC10}. Qubits with linear connectivity need additional SWAP gates for such a realization. In \cite{Gwinner20}, Gwinner et al. provide a way to realize the three-qubit controlled phase gate $\Lambda_{2}(Z)$ via eight CNOT gates on linearly connected qubits, shown below.

\begin{equation}\small
\Qcircuit @C=0.7em @R=0.7em {
& \ctrl{1} & \qw & &&& \gate{T^\dag} & \ctrl{1} & \qw & \ctrl{1} & \qw & \ctrl{1} & \qw & \qw & \ctrl{1} & \qw & \qw  \\
& \ctrl{1} & \qw &&& = \hspace{0.7cm} & \gate{T^\dag} & \targ &  \ctrl{1} & \targ & \ctrl{1} & \targ & \gate{T} & \ctrl{1} & \targ & \ctrl{1} & \qw \\
&  \control \qw& \qw &&&& \gate{T^\dag} & \qw & \targ & \gate{T^\dag} & \targ & \gate{T} & \qw & \targ & \gate{T} & \targ & \qw \\
}
\end{equation}
Here, $T$ is $\pi/8$ gate given by $T = \text{diag}\{1,e^{i\pi/4}\}$ with $i=\sqrt{-1}$.

The four-qubit gate $\Lambda_{3}(Z)$ can be realized by three $\Lambda_{2}(Z)$ gates in a V-shape design (with one clean ancillary qubit) \cite{BBCDMSSSW95}. As pointed in \cite{Maslov16}, we can take advantages of the relative-phase Toffoli gate to reduce the resources of $n$-qubit Toffoli gate constructions. The three-qubit controlled $Y$ gate ($Y=ZX$) can be realized as \cite{Song04}
\begin{equation}\small
\label{eqa:CCY}
\Qcircuit @C=0.7em @R=1.3em {
& \ctrl{1} & \qw &&&& \qw  & \qw  & \qw & \ctrl{2} & \qw & \qw & \qw & \ctrl{2} & \qw \\
& \ctrl{1} & \qw &&& = \hspace{0.7cm} & \qw & \ctrl{1} & \qw & \qw & \qw & \ctrl{1} & \qw & \qw & \qw \\
& \gate{Y} \qw& \qw &&&&  \gate{G} & \targ & \gate{G} & \targ & \gate{G^\dag} & \targ & \gate{G^\dag} & \control \qw & \qw \\
}
\vspace{0.2cm}
\end{equation}
with the $y$-axis rotation gate $G = R_y(\pi/4)$. Then the four-qubit controlled phase gate $\Lambda_3(Z)$ can be constructed as
\begin{equation}\small
\Qcircuit @C=0.7em @R=1.2em {
& \ctrl{1} & \qw &&&&&&& \ctrl{1} & \qw & \ctrl{1} & \qw \\
& \ctrl{2} & \qw &&&&&&& \ctrl{1} & \qw & \ctrl{1} & \qw \\
& \qw \qw& \qw &&& =\hspace{0.5cm} & \hspace{0.4cm} |0\rangle &&& \gate{Y} & \ctrl{1} & \gate{Y^\dag} & \qw \\
& \ctrl{1} & \qw &&&&&&& \qw & \ctrl{1} & \qw & \qw \\
& \control \qw & \qw &&&&&&&  \qw & \control \qw & \qw & \qw   
}
\end{equation}
with one ancillary qubit $|0\rangle$. The rightmost CZ gate in the circuit (\ref{eqa:CCY}) can be cancelled with the CZ gate in $\Lambda_2(Y^\dag)$ when the above $\Lambda_3(Z)$ gate is realized. Note that the CZ gate commutes with the $\Lambda_2(Z)$ gate.

Similar to the realization of $\Lambda_{3}(Z)$ gate, we design the five-qubit controlled phase gate $\Lambda_4(Z)$ via two four-qubit controlled-$Y$ gates $\Lambda_3(Y)$ and one three-qubit controlled phase gate $\Lambda_2(Z)$, in addition to one clean ancillary qubit. The four-qubit controlled gates $\Lambda_3(Y)$ can be decomposed as \cite{Maslov16}
\begin{equation}\small
\Qcircuit @C=0.8em @R=1.2em {
& \ctrl{1} & \qw &&&& \qw & \qw & \qw & \qw & \qw & \ctrl{3} & \qw & \qw & \qw & \ctrl{3} & \qw & \qw & \qw & \qw & \qw & \qw & \qw & \qw & \ctrl{1} & \qw \\
& \ctrl{1} & \qw &&&& \qw & \qw & \qw & \qw & \qw & \qw & \qw & \ctrl{2} & \qw & \qw & \qw & \ctrl{2} & \qw & \qw & \qw & \qw & \qw & \qw & \ctrl{1} & \qw \\
& \ctrl{1} & \qw &&& \raisebox{0.4cm}{=} \hspace{0.9cm} & \qw & \qw & \ctrl{1} & \qw & \qw & \qw & \qw & \qw & \qw & \qw & \qw & \qw & \qw & \qw & \qw & \ctrl{1} & \qw & \qw & \ctrlo{1} & \qw \\
& \gate{Y} \qw& \qw &&&& \gate{H} & \gate{T} & \targ & \gate{T^\dag} & \gate{H} & \targ & \gate{T} & \targ & \gate{T^\dag} & \targ & \gate{T} & \targ & \gate{T^\dag} & \gate{H} & \gate{T} & \targ & \gate{T^\dag} & \gate{H} & \gate{-iZ} & \qw \\
}
\end{equation}
Above circuit requires that the target qubit connects all the control qubits. The rightmost four-qubit controlled gate gives a relative phase $\mp i$ to the states $|1100\rangle$ and $|1101\rangle$, respectively. We do not need to physically realize such gate since it would be cancelled with its inverse in the realization of the $\Lambda_4(Z)$ gate, as shown below.
\begin{equation}\small
\Qcircuit @C=0.7em @R=1.2em {
& \ctrl{1} & \qw &&&&&&& \ctrl{1} & \qw & \ctrl{1} & \qw \\
& \ctrl{1} & \qw &&&&&&& \ctrl{1} & \qw & \ctrl{1} & \qw \\
& \ctrl{2} & \qw &&&&&&& \ctrl{1} & \qw & \ctrl{1} & \qw \\
& \qw \qw& \qw &&& \raisebox{0.5cm}{=} \hspace{0.5cm} & \hspace{0.4cm} |0\rangle &&& \gate{Y} & \ctrl{1} & \gate{Y^\dag} & \qw \\
& \ctrl{1} & \qw &&&&&&& \qw & \ctrl{1} & \qw & \qw \\
& \control \qw & \qw &&&&&&&  \qw & \control \qw & \qw & \qw   
}
\end{equation}
We can also design the $\Lambda_4(Z)$ gate via four $\Lambda_2(Y)$ gates and one $\Lambda_2(Z)$ gate, which requires two clean ancillary qubits. Although this realization has fewer CNOT gates, its fidelity is lower. Seven-qubit superposition is more fragile than the superposition on the six qubits. 

\subsection{\label{subsec:setups}Setup}

\begin{figure}[t!]
\centering
\includegraphics[width=\textwidth]{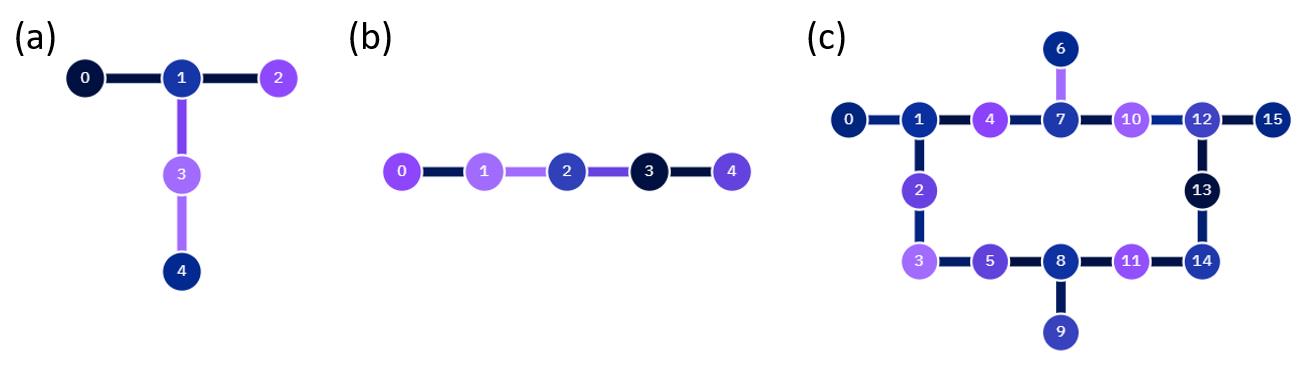}
\caption{Qubits layout of the IBM quantum processors. (a) The five-qubit system, named as Vigo, has the ``T'' connectivity. (b) The five-qubit system, named as Athens, has the linear connectivity. (c) The sixteen-qubit system, is named as Guadalupe. The color on dots represent for the frequency of each qubit. The color on connectivities represent for the error rate of two-qubit gate on the connected two qubits.}
\label{fig_backend}
\end{figure}

We run each circuit with randomly chosen target states in $30$ trials. In each trial, the circuit is run with $8192$ shots to calculate the success probability. We use the same $30$ random target states for different search circuits, in order to compare the results between different search circuits. Since in the unstructured search problem, any target state is equally possible. Random choosen target state is more closed to the practical situation. Besides, state $|1\rangle$ relaxes to the ground state $|0\rangle$ with some probability. Therefore, target strings with larger Hamming weights have lower success probabilities on real devices. Random choosen target state can also mitigate such biases. To make sure that the random choosen target states are not biased, we list those target states in Appendix \ref{App:}.

IBM provides processors with different number of qubits, ranging from one qubit to sixty five qubits. For our purposes, we implement the three-qubit search circuits on the five-qubit processor, Vigo. The four-qubit search circuits are implemented on both Vigo (the five-qubit backend with ``T'' connectivity) and Athens (the five-qubit backend with linear connectivity). The five-qubit search circuits are tested on the sixteen-qubit system, Guadalupe. See Figure \ref{fig_backend} for the topological layout of the qubits in different processors. Note that the connectivity of Vigo is better than the connectivity of Athens. However, Athens has lower gate error rates in average than Vigo. In terms of the metric quantum volume, which characterizes the largest random circuit of equal width
and depth that the computer successfully implements \cite{Cross19}, Athens has quantum volume $32$ while Vigo has $16$.

Besides the success probability of the circuit, we also record the depth of each circuit. The depth is obtained after compiling for each specified backend. The depth obtained in this way represents the real operational length of the circuit. Every quantum processor can only perform four different gates (called the universal gate set) - z-axis rotation gate, X gate, square root X gate and CNOT gate. Oracles encoded different target states have slightly different depths. Combining the success probability and the circuit depth, we calculate the expected depth according to Eqs.~(\ref{def:d_exp}) and (\ref{def:d_exp2}), for the single- and two-stage circuits respectively. 

Recently, Wang et al. introduced the selectivity parameter, $S$, to quantify how distinct the signal is compared to the next most probable outcome \cite{Wang20}. The selectivity is also separately proposed in \cite{Tannu19}, called inference strength. It is important for understanding the quality of results obtained by a search algorithm on real devices. For our purposes, we simply treat it as the ratio between $P_t$ (the probability of obtaining the target string) and the maximum of $P_{nt}$ (the probability of obtaining the non-target string), defined by
\begin{equation}
    S = \frac{P_t}{\max\{P_{nt}\}}.
\end{equation}
Selectivity less than 1 suggests the failure of the implementation. Note that the amplitudes of non-target states are never amplified in the quantum search algorithms. For the classical-quantum hybrid circuit, we only consider the selectivity for the results obtained from the quantum algorithm. In other words, $P_t$ is the probability of obtaining the target substring. For the multi-stage circuits, we choose the minimal selectivity among the circuits from the different stages. Note that the selectivity reveals the probabilistic distribution of the implementation results, rather than a theoretical parameter. 

We also benchmark the results via the degraded ratio, defined as
\begin{equation}
\label{def:degraded}
    R_\text{IBMq} = \frac{P_\text{IBMq}}{P_\text{theo}}.
\end{equation}
Here, the probability $P_\text{theo}$ is the theoretical success probability of finding the target string (the theoretical success probability of the search circuit); probability $P_\text{IBMq}$ is the success probability obtained from the IBM quantum processors. Degraded ratio has been reported to decay exponentially with the number of two-qubit gates in \cite{Gwinner20}. Such fast degradation is the motivation for our multi-stage strategy.  

For different circuits, we use the following notations. The Grover operator with the diffusion operator $D_m$ is denoted as Dm. Measurement on $p$ qubits is Mp. If there is a classical initialization on q number of qubits, then it is Gq (see the hybrid search algorithm in Sec.~\ref{subsec:hybrid_search}). Note that the italic notation $G_m$ is for the Grover operator defined in Eq.~(\ref{def G m}). The left to right ordering represents the order in which these operations are carried out. We always specify the search domain for each circuit notations. For example, the three-qubit search circuit G1D2M2 represents the random initialization of one qubit followed by the Grover operator with a two-qubit diffusion operator and then measure these two qubits.

For the two-stage algorithm, we follow the same rules, but each of the stages is separated by a vertical line ``$|$''. For example, the three-qubit search circuit D2M1$|$D2M2 is a two-stage algorithm. In the first stage, the Grover operator $G_2$ (with two-qubit diffusion operator) is applied, then one of the qubits is measured (acted upon by the diffusion operator). The second stage is to initialize the state according to the results from the first stage (see Sec.~\ref{subsec:divde_conquery} for details), then apply the Grover operator $G_2$ (with two-qubit diffusion operator) followed by the measurement on the two qubits (to be searched). A list of explanations on all the circuit notations can be found in Appendix. See Table \ref{Table:3q_names} for the three-qubit search circuits  and Table \ref{Table:4q_names} for the four-qubit search circuits.

\subsection{\label{subsec:3q}Three-qubit cases}

\begin{figure*}[t!]
\includegraphics[width=\textwidth]{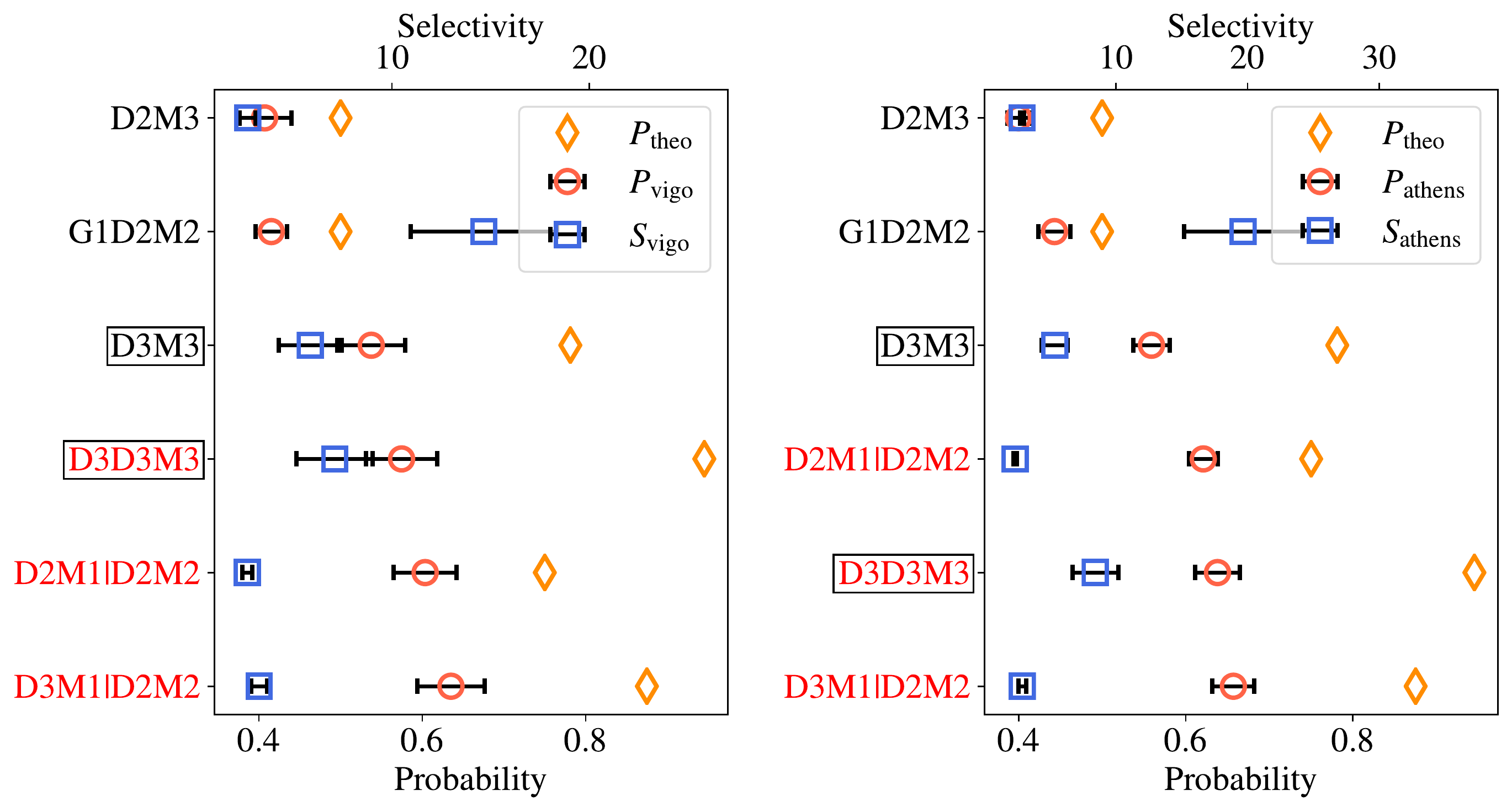}
\caption{Success probabilities of the three-qubit search circuits on the IBM quantum processors Vigo (left) and Athens (right). Circuit names with black and red colors are the circuits with one and two oracles, respectively. The standard Grover's algorithm circuits are boxed. Circuits are ordered as the magnitude of the success probabilities obtained from the Vigo machine (red circle). Standard deviations are obtained from $30$ trials with random target states. Each trial has $8192$ shots.}
\label{fig_3q_prob}
\end{figure*}

\begin{figure*}[t!]
\includegraphics[width=\textwidth]{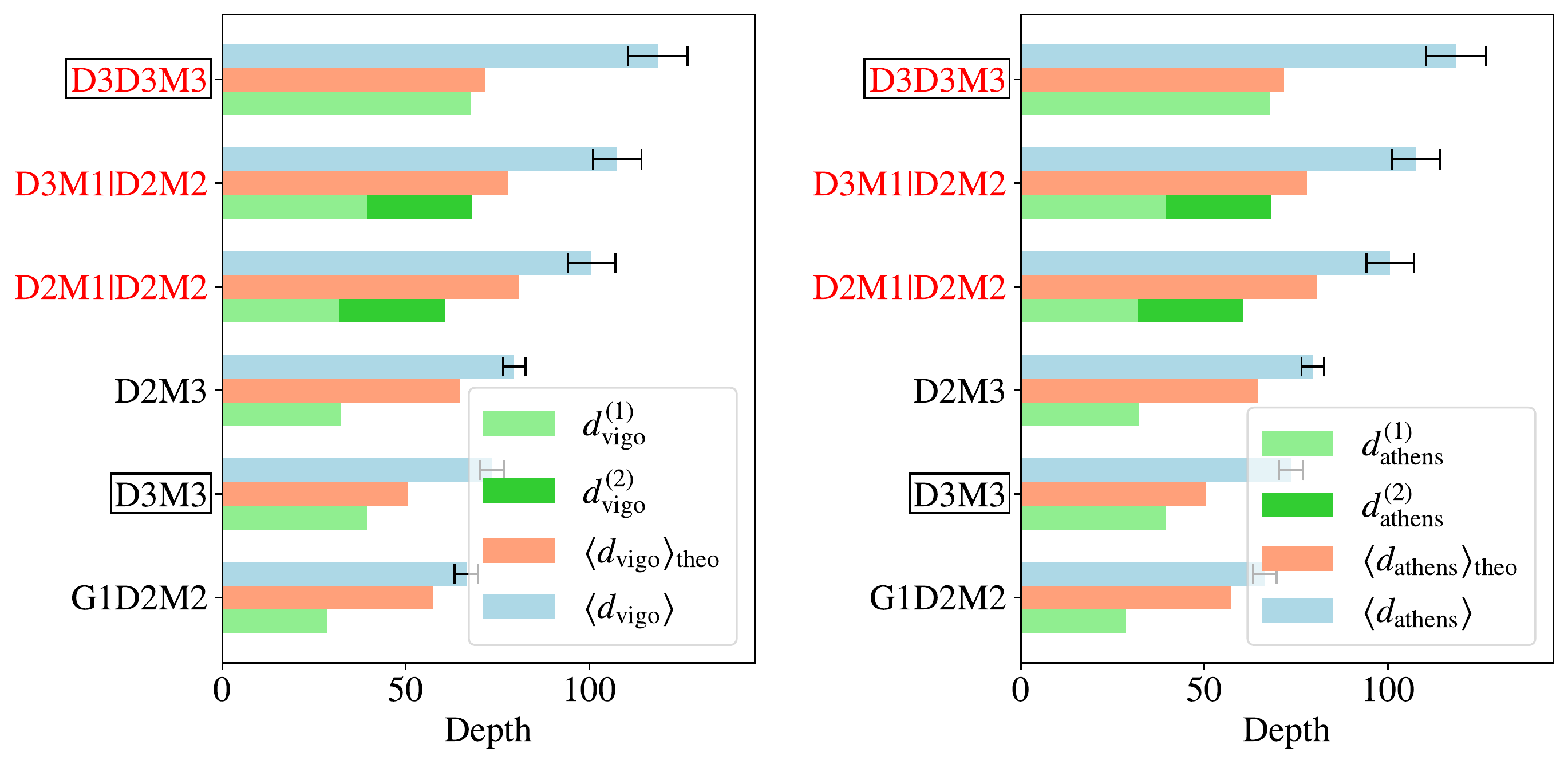}
\caption{Circuit depths and expected depths of the three-qubit search circuits on the IBM quantum processors Vigo (left) and Athens (right). Here $d^{(1)}$ or $d^{(2)}$ is the depth of first or second stage circuit; $\langle d_\text{vigo}\rangle_\text{theo}$ or $\langle d_\text{athens}\rangle_\text{theo}$ is the theoretical expected depth of the circuit (given by the theoretical success probability) on Vigo or Athens; $\langle d_\text{vigo}\rangle$ or $\langle d_\text{athens}\rangle$is the real expected depth of the circuit (given by the real success probability) on Vigo or Athens. Circuit names with black and red colors indicate the circuits with one and two oracles, respectively. The standard Grover's algorithm circuits are boxed. Circuits are ordered according to the magnitude of the expected depth obtained from the Vigo machine (top blue bar). Standard deviations are obtained from $30$ trials with random target states. Each trial has $8192$ shots.}
\label{fig_3q_depth}
\end{figure*}

Three-qubit Grover's algorithm gives the theoretical success probabilities of $0.781$ and $0.945$ with one and two Grover iterations, respectively. These two circuits are denoted as D3M3 and D3D3M3. We design another two circuits with one oracle, i.e., D2M3 and G1D2M2, which both give the $0.5$ theoretical success probabilities. We design two different two-stage three-qubit search circuits. They both find one bit of the target in the first stage. One uses the three-qubit diffusion operator, i.e., D3M1, the other uses the two-qubit diffusion operator, i.e., D2M1. In the second stage, the circuits are equivalent to the two-qubit search, which gives success probability as $1$ with just a single Grover iteration, i.e., D2M2. The detailed explanations about these six circuits can be found in Appendix (Table \ref{Table:3q_names}).

We plot the success probabilities, as well as the selectivities, of the three-qubit search circuits in Figure~\ref{fig_3q_prob}. The detailed data, as well as the thirty random target states, can be found in Appendix \ref{App:}. To exclude any possible biases from the random target states, we also provide supplementary data based on the average target states in Appendix \ref{App:B} (Table \ref{Table:3q_results_average}). Both on the Vigo and Athens machines, the two-stage circuit D3M1$|$D2M2 gives the largest success probabilities. Both the two-stage circuits have higher success probabilities than Grover's realization D3D3M3, though D3D3M3 has the largest theoretical success probability. Such results demonstrate the significance of the divide-and-conquer strategy. The circuit G1D2M2 has the largest selectivity, which shows its robustness against the errors. The circuits D2M3 and G1D2M2 have the identical implementations except for the initial states. The success probability of G1D2M2 is slightly higher than D2M3 because the latter is only manipulating the two-qubit superposition states. 

Incorporating the circuit depth with their success probabilities, we plot the expected depth in Figure~\ref{fig_3q_depth}. There are three different categories of the depth parameter. The first is the depth of the circuit, such as $d(\text{D3M3})$. The second is the theoretical expected depth, given by the theoretical success probability of the circuits. The last is the expected depth on real machines, given by the success probabilities obtained from the IBM quantum processors. The least expected depth on the Vigo and Athens machines are both given by the circuit G1D2M2. Although G1D2M2 has lower success probability than D3M3, its shallow depth can find the target state more efficiently. The two two-stage circuits have lower expected depth than Grover's D3D3M3, since they have lower depth realizations while maintaining higher success probabilities. Both D3M3 and D3D3M3 are the standard Grover's algorithm. Neither the one-oracle D3M3 nor the two-oracle D3D3M3 gives the optimal expected depth. Depth optimizations for the search algorithm are necessary when running on the real quantum devices.   

\begin{figure*}[htbp]
\centering
    \includegraphics[width=\textwidth]{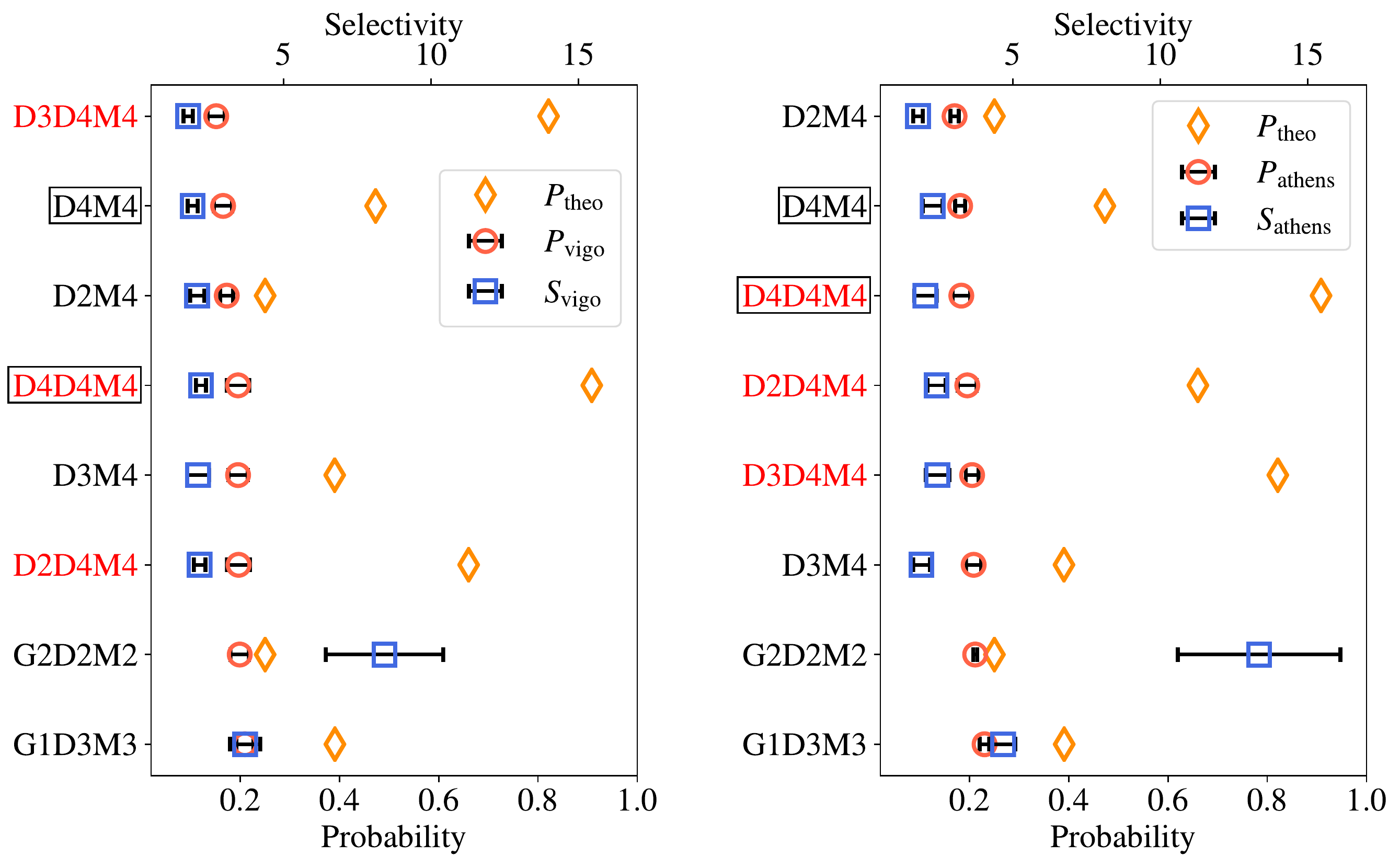}
    \caption{Probabilities of the one-stage four-qubit search circuits on the IBM quantum processor Vigo (left figure) and Athens (right figure). Circuit names with black and red color are circuits with one and two oracles, respectively. The standard Grover's algorithm circuits are boxed. Circuits are ordered according to the magnitude of the success probabilities (red circle). Standard deviations are obtained from $30$ trials with random target states. Each trial has $8192$ shots.}   
\label{fig_4q_prob_one}

\bigskip

\includegraphics[width=\textwidth]{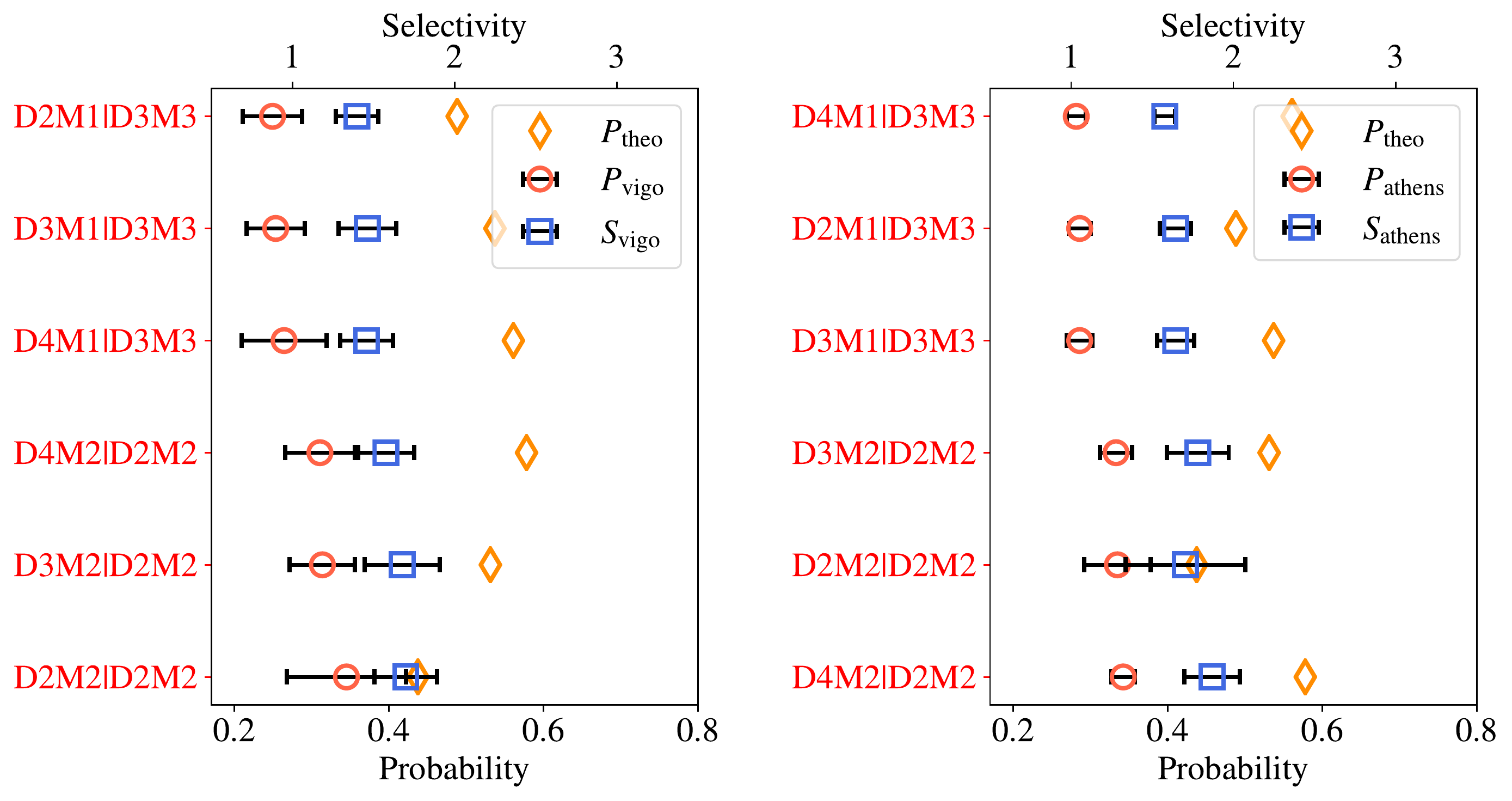}
    \caption{Probabilities of the two-stage four-qubit search circuits on the IBM quantum processor Vigo (left figure) and Athens (right figure). Circuits are ordered according to the magnitude of the success probabilities (red circle). Standard deviations are obtained from $30$ trials with random target states. Each trial has $8192$ shots.}   
\label{fig_4q_prob_two}
\end{figure*}

\begin{figure*}[htbp]
\centering
    \includegraphics[width=\textwidth]{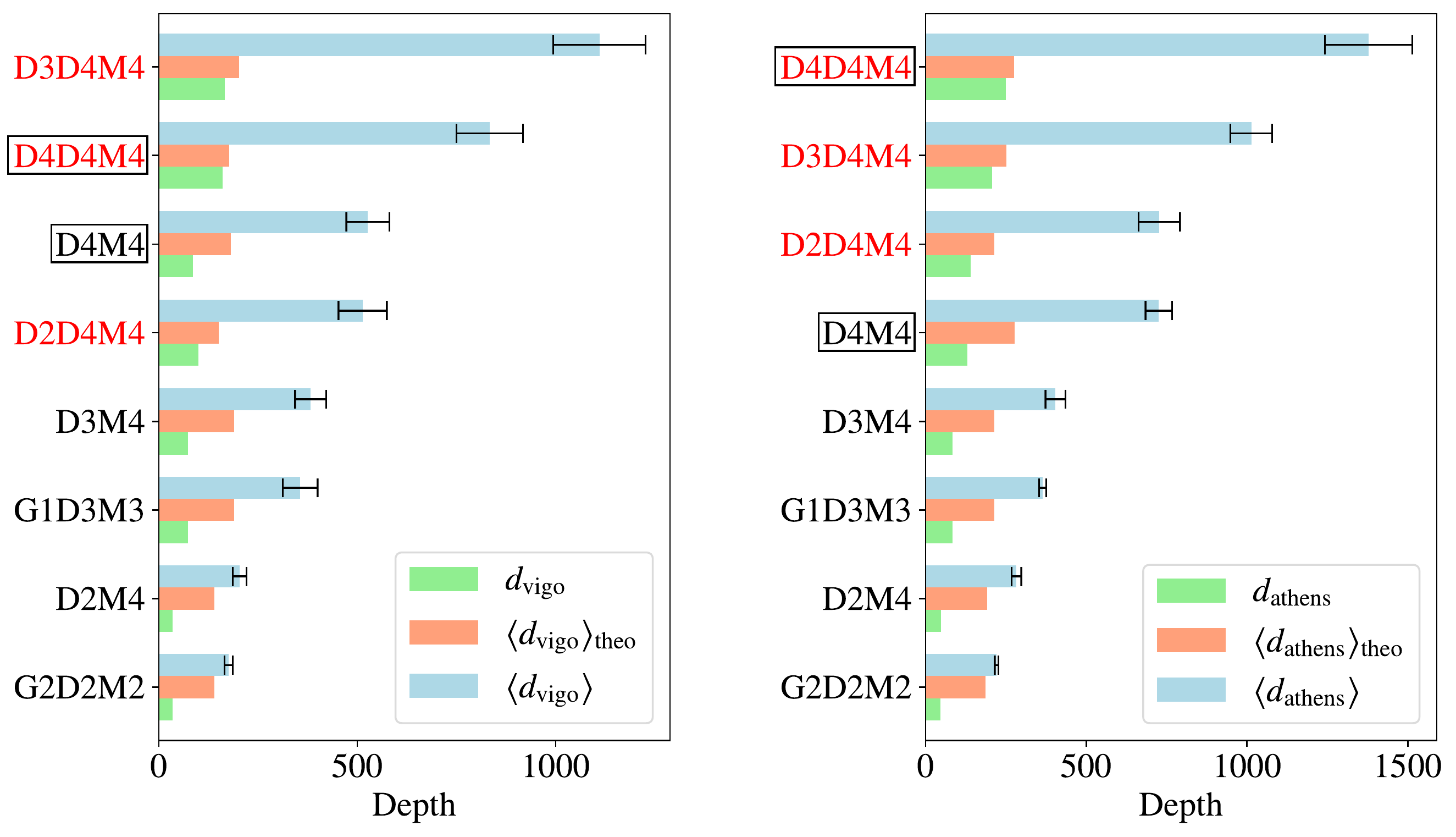}
    \caption{Circuit depths and expected depths of the one-stage four-qubit search circuits on the IBM quantum processor Vigo (left figure) and Athens (right figure). Here $d_\text{vigo}$ or $d_\text{athens}$ is the circuit depth on Vigo or Athens; $\langle d_\text{vigo}\rangle_\text{theo}$ or $\langle d_\text{athens}\rangle_\text{theo}$ is the theoretical expected depth of the circuit (given by the theoretical success probability) on Vigo or Athens; $\langle d_\text{vigo}\rangle$ or $\langle d_\text{athens}\rangle$is the real expected depth of the circuit (given by the real success probability) on Vigo or Athens. Circuit names with black and red colors are circuits with one and two oracles, respectively. The standard Grover's algorithm circuits are boxed. Circuits are ordered according to the magnitude of the expected depth (top blue bar). Standard deviations are obtained from $30$ trials with random target states. Each trial has $8192$ shots.}   
\label{fig_4q_depth_one}

\bigskip

\includegraphics[width=\textwidth]{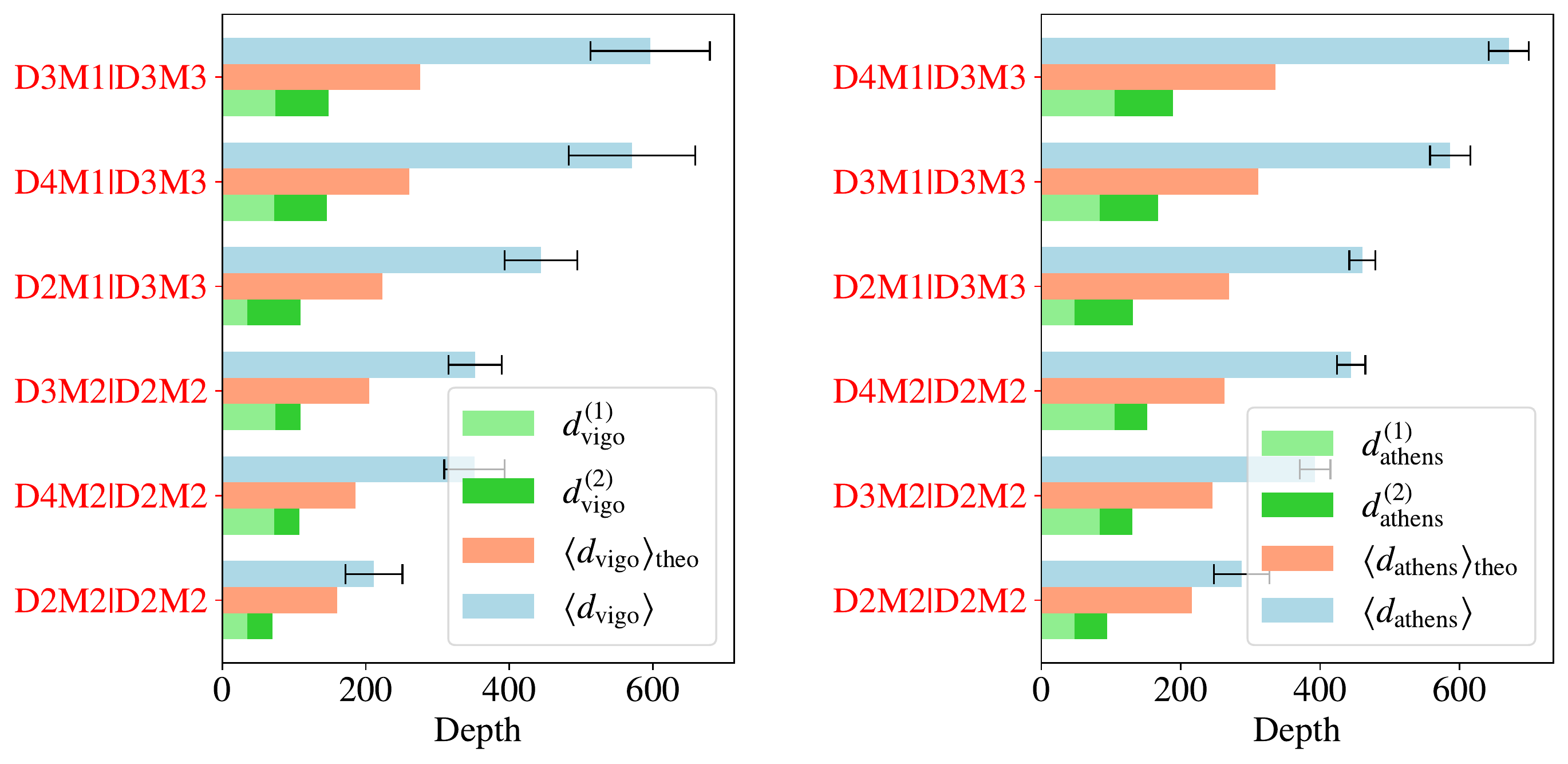}
    \caption{Circuit depths and expected depths of the two-stage four-qubit search circuits on the IBM quantum processor Vigo (left figure) and Athens (right figure). Here $d^{(1)}$ or $d^{(2)}$ is the depth of first or second stage circuit; $\langle d_\text{vigo}\rangle_\text{theo}$ or $\langle d_\text{athens}\rangle_\text{theo}$ is the theoretical expected depth of the circuit (given by the theoretical success probability) on Vigo or Athens; $\langle d_\text{vigo}\rangle$ or $\langle d_\text{athens}\rangle$is the real expected depth of the circuit (given by the real success probability) on Vigo or Athens. Circuits are ordered according to the magnitude of the expected depth (top blue bar). Standard deviations are obtained from $30$ trials with random target states. Each trial has $8192$ shots.} 
\label{fig_4q_depth_two}
\end{figure*}

\subsection{\label{subsec:4q}Four-qubit cases}

Four-qubit search has more scope for exploiting the partial diffusion operators than the three-qubit search. Including the standard Grover's algorithm with one and two oracles, we design a total of fourteen different circuits and test them on both the Vigo and Athens processors. See Appendix (Table \ref{Table:4q_names}) for detailed explanations on each circuit. Among the fourteen circuits, eight are one-stage circuits and six are two-stage circuits. Success probabilities of the one-stage circuits are plotted in Figure~\ref{fig_4q_prob_one}. For the two-stage circuits, the success probabilities are plotted in Figure~\ref{fig_4q_prob_two}. Also see Table \ref{Table:random_target} in Appendix \ref{App:} for the thirty random target states. The supplementary data on the average target states can be found in Appendix \ref{App:B} (Table \ref{Table:4q_results_average}).

Among the eight one-stage search circuits, D4D4M4 gives the largest theoretical probability $0.908$. However, the real success probability is degraded below to $0.2$ (both on Vigo and Athens) due to the actual implementation that has longer depth. On Vigo and Athens both, G1D3M3 gives the largest success probability among the one-stage circuits on real machines. On the Vigo machine, the two-oracle circuit D2D4M4 has larger success probability than the Grover's two-oracle circuit D4D4M4. Similar results are found on Athens. Local diffusion operators may decrease the theoretical success probability, but its low depth overcomes such disadvantages on quantum devices. 

For more information on the two-stage circuits, see Figure \ref{fig_4q_prob_two}. There are two types of two-stage four-qubit circuits. One type finds a target bit at the first stage; then finds the rest of the three target bits in the second stage. The other type searches two target bits in each of the two stages, which provides a higher probability than the first dividing strategy. Circuits implemented on Vigo have quite similar results on Athens. The two-stage circuit, D2M2$|$D2M2, gives the average success probability of $0.345$. The two-oracle Grover's search, D4D4M4, has the average success probability of $0.195$. Notably, our divide-and-conquer circuit D2M2$|$D2M2 nearly doubles the success probability of the standard Grover's circuit D4D4M4. Compared to a recent study \cite{Gwinner20}, our success probability for the four-qubit search algorithm exceeds $0.3$, which is significant.

For the depth (the circuit depth, the theoretical expected depth and the expected depth on the real devices) of the one-stage and the two-stage four-qubit search circuits, see Figures~\ref{fig_4q_depth_one} and \ref{fig_4q_depth_two}. Note that because of the better connectivity of the Vigo, every circuit has slightly longer depth on Athens as compared to Vigo. The averaged circuit depth of G2D2M2 is less than the half of one-oracle Grover's circuit D4M4. However, the overall success probability of G2D2M2 is higher than D4M4. The expected depth of circuit G2D2M2 is far shorter than that of D4M4. The circuit G2D2M2 has the shortest theoretical expected depth both on the Vigo and Athens.

In the two-stage circuits, the two-two dividing has shorter depth than the one-three dividing. The two-two dividing exploits the local diffusion operator more efficiently. Note that the one-qubit quantum search algorithm is not well defined therefore we do not have the three-one dividing. The expected depth of two-oracle Grover's circuit D4D4M4 is $> 800$ ($833.99$ on Vigo and $1378.31$ on Athens). The divide-and-conquer circuit (D2M2$|$D2M2) suppresses the expected depth below $300$ ($211.35$ on Vigo and $287.58$ on Athens).

\begin{figure}[t!]
    \centering
    \includegraphics[width=0.5\textwidth]{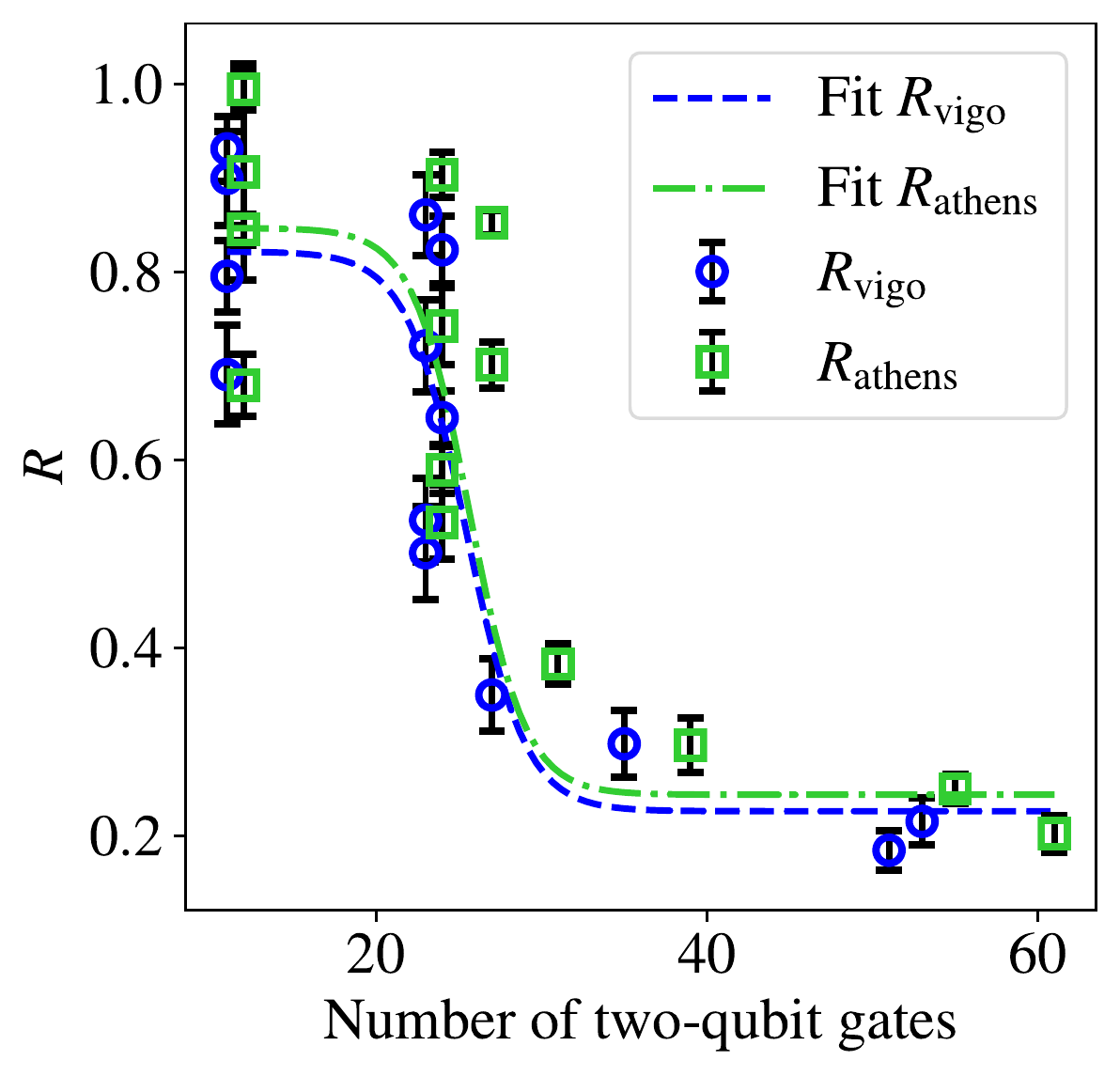}
    \caption{Degraded ratio of the success probabilities vs. the number of two-qubit gates. The curve is fitted according to the logistic function. If the number of two-qubit gates exceeds thirty, the noise may lead to an inefficient search.}
\label{fig_R}
\end{figure}

Quantum algorithms with long circuits can not be directly divided into several stages, with new initialized states during the algorithm. Measurements in the middle will wipe out the established coherence between qubits. Suppose that the long circuit has the success probability of $p$. Dividing it into two parts gives the success probabilities $p_1$ and $p_2$ for each part, respectively. Previous theoretical study shows that $p>p_1p_2$ \cite{Zhang20}. However, our tests demonstrate that $p'<p'_1p'_2$ on real quantum computers. The explanation relies on benchmarking the parameter, the degraded ratio $R$, defined in Eq.~(\ref{def:degraded}). Based on our data from the four-qubit search circuits, we plot the degraded ratio versus the number of two-qubit gates in Figure \ref{fig_R}. The degraded ratio drops dramatically when the circuits have more than $30$ two-qubit gates. In other words, if the number of two-qubit gates exceeds thirty, the noises may surpass the amplified amplitude of the target state. The data is fitted by the logistic function
\begin{equation}
    R(x) = \frac{a}{1+\exp(-b(x-c))}+d.
\end{equation}
Here, $x$ is the number of two-qubit gates and the numbers $\{a,b,c,d\}$ are the parameters to be fitted. Similar benchmarking results have been reported in \cite{Gwinner20}. This exponential drop is the motivation for our divide-and-conquer designs and explains why the divide-and-conquer circuits have better performance than the one-stage circuits.

\subsection{\label{subsec:5q}Five-qubit cases}

\begin{table}[t!]
	\caption{Results on the five-qubit search circuits. Circuit names with $\dag$ are the standard Grover's algorithms. Standard deviations are obtained from $30 \times 8192 $ shots with random target states.}
	\label{Table:5q_results}
\begin{tabular}{ccccccc}\hline
		Circuits & $P_\text{theo}$ & $P_\text{guada}$ & $S_\text{guada}$ & $d_\text{guada}$ & $\langle d_\text{guada}\rangle_\text{theo}$ & $\langle d_\text{guada}\rangle$  \\ \hline
        D5M5$^\dag$ & 0.258 & $0.0257\pm 0.0038$ & $0.363\pm0.076$ & 122.37 & 474.30 & $4856.75\pm675.33$\\
        G2D3M3 & 0.195 & $0.0654\pm 0.0087$ & $1.880\pm0.355$ & 82.00 & 420.51 & $1276.61\pm174.98$ \\
        G3D2M2 & 0.125 & $0.0963\pm0.0021 $ & $8.282\pm1.039$ & 53.53 & 428.24 & $556.02\pm24.67$\\
        D2M2$\mid$D3M3 & 0.268 & $0.0667\pm0.0150$ & $0.904\pm0.215$ & 135.53 & 505.71 & $2121.44\pm434.46$ \\
        D3M3$\mid$D2M2 & 0.289 & $0.1014\pm0.0283$ &  $0.764\pm0.313$ & 135.67 & 469.45 & $1444.97\pm395.33$\\\hline
	\end{tabular}
\end{table} 

To the best of our knowledge, the five-qubit search algorithm has never been successfully implemented on any of the IBM quantum processors. The five-qubit controlled Toffoli gate $\Lambda_4(X)$ requires more gates as well as more connectivity between qubits. We test the standard one-oracle Grover's algorithm D5M5 on the sixteen-qubit processor Guadalupe (only six qubits are used by our circuits). We choose the six qubits with least two-qubit gate errors. The probability of finding the target string is degraded to a lower value than the probability of finding the non-target string. See the results listed in Table \ref{Table:5q_results}. The selectivity is below 1, which implies the failure of the algorithm. Classical unstructured search with one oracle can find the target string with the probability 0.0625: randomly pickup a string then verify it with the oracle; if it is not the target string then randomly pickup another one ($\frac {1} {2^5}+\frac{2^5-1}{2^5}\times\frac{1}{2^5-1}=0.0625$). The quantum search D5M5 (with the averaged success probability of $0.0257$) is worse than the classical case. Similar results on the average target states can be found in Table \ref{Table:5q_results_average} in Appendix \ref{App:B}.

Although there are plenty of implementations of the five-qubit one-oracle search via the local diffusion operators, most of them give failed results (selectivity less than 1 and success probability less than classical search). However, we found that the classical-quantum hybrid search circuits G2D3M3 and G3D2M2 have selectivities larger than 1, see Table \ref{Table:5q_results}. G2D3M3 gives success probability higher than random pick, but relatively same as the classical search. G3D2M2 gives the average success probability $0.0963$, which is higher than the classical algorithm (with one oracle). G3D2M2 randomly chooses three bits to perform the normalized two-qubit search algorithm. Since there are more two-qubit gates acting on the two qubits to be searched, we choose the physical two qubits with least errors of two-qubit gates. In our setup, physical qubits 13 and 14 are choosen, see Fig. \ref{fig_backend}. The theoretical success probability is $1/8=0.125$, since the two-qubit search with one oracle has $1$ as the success probability. Theoretically, G3D2M2 has only half success probability compared to D5M5. In practical, the success probability of G3D2M2 is three times higher than Grover's algorithm D5M5 because of its shallow depth. In the three- and four-qubit cases, such hybrid classical-quantum circuits always have the largest selectivity (see Figures~\ref{fig_3q_prob} and \ref{fig_4q_prob_one}). Thus, we can expect that the five-qubit search G3D2M2 stands out in the five-qubit search implementations. 

Based on the success of G2D3M3 and G3D2M2, we test the divide-and-conquer circuits D2M2$\mid$D3M3 and D3M3$\mid$D2M2, which replace the random guesses in G2D3M3 and G3D2M2 by the quantum partial search algorithms. Although we have selectivities of G2D3M3 and G3D2M2 larger than 1, the selectivities D2M2$\mid$D3M3 and D3M3$\mid$D2M2 are less than 1. It suggests that the quantum partial search algorithm in the first stage is not efficient, which does not provide quantum speedup. 


%
\section{\label{sec:conclusion}Conclusion}

In this paper, we have implemented the three-, four- and five-qubit search algorithms on the IBM quantum processors. Grover’s algorithm does not provide the optimal performance on the NISQ devices. To reduce the noise, we have designed the quantum search circuits using the local diffusion operators. There are three different strategies to exploit the local diffusion operators. We realized the four-qubit search algorithm with the highest success probability compared to other studies. We also successfully ran the five-qubit search algorithm on the IBM quantum devices for the first time. Additionally, the use of multi-stage circuits makes it possible to run the search in parallel. We envision our work still be useful in post-NISQ era, since the lower depth circuits would require less resources for the error corrections. 

\begin{acknowledgements}

This research used resources of the Oak Ridge Leadership Computing Facility, which is a DOE Office of Science User Facility supported under Contract DE-AC05-00OR22725. This material is based upon work supported by the U.S. Department of Energy, Office of Science, National Quantum Information Science Research Centers, Co-design Center for Quantum Advantage (C2QA) under contract number DE-SC0012704. P. R. participated in a program hosted by the Mathematical Sciences Research Institute in Berkeley, California, during the Spring 2021 semester, which is supported by the National Science Foundation under Grant No. DMS-1928930.
	
\end{acknowledgements}

\appendix

\section{\label{App:} Supplementary data for random target states}

Calibration parameters of the IBM quantum processors (for the random target states) are listed in Table \ref{Table:machines}. The calibration data was retrieved at the time when the circuits were implemented. The thirty random choosen target states for the three-, four-, and five-qubit search are listed in Table \ref{Table:random_target}. The explanations on the three-qubit circuits are listed in Table \ref{Table:3q_names}. The experimental results for the three-qubit circuits are presented in Table \ref{Table:3q_results}. The naming convention for the four-qubit circuits are listed in Table \ref{Table:4q_names}. The theoretical and experimental data for the four-qubit search circuits are presented in Tables \ref{Table:4q_parameters} and \ref{Table:4q_results} respectively. 

\begin{table}[ht]
	\caption{Calibration specs for the IBM quantum processors were retrieved on the day of data (for random target states)}.
	\label{Table:machines}
\begin{tabular}{cccccccc}\hline
		Backends & CNOT error & Readout error & $T_1$ & $T_2$ & Quantum volume & Version & Date  \\ \hline
        Vigo & 8.627e-3 & 3.222e-2 & 98.13 & 66.88 & 16 & 1.3.5 & Jan. 9, 2021 \\
        Athens & 9.262e-3 & 1.842e-2 & 84.55 & 102.56 & 32 & 1.3.6 & Jan. 17, 2021 \\
         Guadalupe & 1.245e-2 & 2.056e-2 & 78.15 & 90.72 & 32 & 1.2.18 & June 12, 2021 \\\hline
	\end{tabular}
\end{table}

\begin{table}[ht]
	\caption{Thirty random generated target states for the three-, four-, and five-qubit search algorithms.}
	\label{Table:random_target}
\begin{tabular}{ccccccccccc}\hline
        Number of qubits & \multicolumn{10}{c}{Thirty random target states} \\ \hline
		 & 001 & 101 & 010 & 001 & 001 & 111 & 010 & 111 & 001 & 100 \\
		Three & 011 & 000 & 100 & 111 & 010 & 011 & 110 & 111 & 110 & 011  \\
		& 101 & 111 & 110 & 001 & 001 & 000 & 001 & 001 & 001 & 001  \\ \hline
		& 1001 & 1101 & 1010 & 0001 & 1110 & 0010 & 1001 & 0100 & 0011 & 0111 \\
		Four & 0001 & 0101 & 1110 & 0000 & 1010 & 1010 & 0101 & 0011 & 0001 & 0000 \\
		& 1100 & 0110 & 1111 & 0111 & 0000 & 0101 & 1101 & 1111 & 1000 & 0111 \\\hline
		& 01010 & 10001 & 01011 & 01000 & 11111 & 00000 & 00000 & 00100 & 01010 & 00010 \\
		Five & 01011 & 11100 & 10101 & 11010 & 00100 & 10100 & 01010 & 11001 & 01100 & 10001 \\
		& 00011 & 01101 & 00011 & 10000 & 10100 & 10000 & 11000 & 10100 & 11111 & 11000 \\\hline
	\end{tabular}
\end{table} 

\begin{table}[ht]
	\caption{Naming explanations for the three-qubit search circuits. Suppose that the target string is $t_1t_2t_3$, corresponding to the target state $|t_1t_2t_3\rangle$. Circuit names with $\dag$ are the standard Grover's algorithms. }
	\label{Table:3q_names}
\begin{tabular}{cccccc}\hline
		Circuits & Oracle numbers & Initial state & Operation & Measurement & Remarks on the second stage \\ \hline
        D3M3$^\dag$ & 1 & $H^{\otimes 3}|0\rangle^{\otimes 3}$ & $G_3$ &   All qubits & NA \\
        D2M3 & 1 & $H^{\otimes 3}|0\rangle^{\otimes 3}$ & $G_2$ &   All qubits & NA \\
        G1D2M2 & 1 & $|t_1\rangle\otimes H^{\otimes 2}|0\rangle^{\otimes 2}$ & $G_2$ &  Qubits of $|t_2t_3\rangle$ & NA \\
        D3D3M3$^\dag$ & 2 & $H^{\otimes 3}|0\rangle^{\otimes 3}$ & $G^2_3$ &   All qubits & NA \\ 
        D3M1$\mid$D2M2 & 2 & $H^{\otimes 3}|0\rangle^{\otimes 3}$ & $G_3$ & Qubit of $|t_1\rangle$ & Equivalent to G1D2M2 \\
        D2M1$\mid$D2M2 & 2 & $H^{\otimes 3}|0\rangle^{\otimes 3}$ & $G_2$ & Qubit of $|t_1\rangle$ & Equivalent to G1D2M2 \\\hline
	\end{tabular}
\end{table}

\begin{table}[ht]
	\caption{Parameters of the three-qubit search circuits on the Vigo and Athens. Circuit names with $\dag$ are the standard Grover's algorithms.}
	\label{Table:3q_theo}
\begin{tabular}{ccccccc}\hline
		Circuits & $P_\text{theo}$ & $d^{(1)}_\text{vigo}$, $d^{(2)}_\text{vigo}$ & $\langle d_\text{vigo}\rangle_\text{theo}$  & $d^{(1)}_\text{athens}$, $d^{(2)}_\text{athens}$ & $\langle d_\text{athens}\rangle_\text{theo}$ \\ \hline
        D3M3$^\dag$ & 0.781  & 39.43 & 50.47 & 40.03 & 51.24 \\
        D2M3 & 0.5 & 32.30 & 64.60 & 33.17 & 66.33 \\
        G1D2M2 & 0.5 & 28.70 & 57.40 & 29.73 & 59.47 \\
        D3D3M3$^\dag$ & 0.945 & 67.83 & 71.76 & 68.63 & 72.60 \\ 
        D3M1$\mid$D2M2 & 0.875 & $39.43$, $28.70$ & 79.73 & $40.03$, $29.73$ & 79.73 \\
        D2M1$\mid$D2M2 & 0.750 & $31.87$, $28.70$ & 83.87 & $33.17$, $29.73$ & 83.87\\\hline
	\end{tabular}
\end{table} 

\begin{table}[ht]
	\caption{Success probabilities, selectivities and depths of the three-qubit search circuits. Circuit names with $\dag$ are the standard Grover's algorithms. Standard deviations are obtained from $30$ trials with random target states. Each trial has $8192$ shots.}
	\label{Table:3q_results}
\begin{tabular}{ccccccc}\hline
		Circuits & $P_\text{vigo}$ & $S_\text{vigo}$ & $\langle d_\text{vigo}\rangle$ & $P_\text{athens}$ & $S_\text{athens}$ & $\langle d_\text{athens}\rangle$ \\ \hline
        D3M3$^\dag$ & $0.538\pm 0.042$ & $5.86\pm1.59$ & $73.58\pm3.29$ & $0.559\pm 0.022$ & $5.36\pm0.99$ & $71.71\pm3.29$ \\
        D2M3 & $0.407\pm 0.033$ & $2.70\pm0.40$ & $79.52\pm3.09$ & $0.400\pm0.013$ & $2.88\pm0.15$ & $83.00\pm5.28$ \\
        G1D2M2 & $0.415\pm0.019$ & $14.65\pm3.70$ & $66.49\pm 3.20$ & $0.443\pm0.019$ & $19.66\pm4.49$ & $66.60\pm3.18$ \\
        D3D3M3$^\dag$ & $0.575\pm 0.044$ & $7.10\pm 1.93$ & $118.29\pm8.16$ & $0.638\pm0.027$ & $8.44\pm1.73$ & $107.67\pm4.13$ \\ 
        D3M1$\mid$D2M2 & $0.635\pm0.042$ & $3.28\pm0.40$ & $107.58\pm6.58$ & $0.657\pm0.025$ & $2.89\pm0.30$ & $106.31\pm7.11$\\
        D2M1$\mid$D2M2 & $0.604\pm0.038$ & $2.67\pm0.25$ & $100.61\pm6.44$ & $0.621\pm0.017$ & $2.35\pm0.13$ & $101.26\pm5.87$\\\hline
	\end{tabular}
\end{table}

\begin{table}[ht]
	\caption{Naming explanations for the four-qubit search circuits. Suppose that the target string is $t_1t_2t_3t_4$, corresponding to the target state $|t_1t_2t_3t_4\rangle$. Circuit names with $\dag$ are the standard Grover's algorithms. }
	\label{Table:4q_names}
\begin{tabular}{cccccc}\hline
		Circuits & Oracle numbers & Initial state & Operation & Measurement & Remarks on the second stage \\ \hline
        D4M4$^\dag$ & 1 & $H^{\otimes 4}|0\rangle^{\otimes 4}$ & $G_4$ &   All qubits & NA \\
        D3M4 & 1 & $H^{\otimes 4}|0\rangle^{\otimes 4}$ & $G_3$ &   All qubits & NA \\
        D2M4 & 1 & $H^{\otimes 4}|0\rangle^{\otimes 4}$ & $G_2$ &   All qubits & NA \\
        G1D3M3 & 1 & $|t_1\rangle\otimes H^{\otimes 3}|0\rangle^{\otimes 3}$ & $G_3$ &   Qubits of $|t_2t_3t_4\rangle$ & NA \\
        G2D2M2 & 1 & $|t_1t_2\rangle\otimes H^{\otimes 2}|0\rangle^{\otimes 2}$ & $G_2$ &   Qubits of $|t_3t_4\rangle$ & NA  \\
        D4D4M4$^\dag$ & 2 & $H^{\otimes 4}|0\rangle^{\otimes 4}$ & $G^2_4$ &   All qubits & NA \\
        D3D4M4 & 2 & $H^{\otimes 4}|0\rangle^{\otimes 4}$ & $G_4G_3$ &   All qubits & NA \\
        D2D4M4 & 2 & $H^{\otimes 4}|0\rangle^{\otimes 4}$ & $G_4G_2$ &   All qubits & NA \\
        D4M1$\mid$D3M3 & 2 & $H^{\otimes 4}|0\rangle^{\otimes 4}$ & $G_4$ &   Qubit of $|t_1\rangle $ & Equivalent to G1D3M3 \\
        D3M1$\mid$D3M3 & 2 & $H^{\otimes 4}|0\rangle^{\otimes 4}$ & $G_3$ &   Qubit of $|t_1\rangle $ & Equivalent to G1D3M3 \\
        D2M1$\mid$D3M3 & 2 & $H^{\otimes 4}|0\rangle^{\otimes 4}$ & $G_2$ &   Qubit of $|t_1\rangle $ & Equivalent to G1D3M3 \\
        D4M2$\mid$D2M2 & 2 & $H^{\otimes 4}|0\rangle^{\otimes 4}$ & $G_4$ &   Qubits of $|t_1t_2\rangle $ & Equivalent to G2D2M2 \\
        D3M2$\mid$D2M2 & 2 & $H^{\otimes 4}|0\rangle^{\otimes 4}$ & $G_3$ &   Qubits of $|t_1t_2\rangle $ & Equivalent to G2D2M2 \\
        D2M2$\mid$D2M2 &  2 & $H^{\otimes 4}|0\rangle^{\otimes 4}$ & $G_2$ &   Qubits of $|t_1t_2\rangle $ & Equivalent to G2D2M2 \\\hline
	\end{tabular}
\end{table}

\begin{table}[ht]
	\caption{Parameters of the four-qubit search circuits on the Vigo and Athens. Circuit names with $\dag$ are the standard Grover's algorithms.}
	\label{Table:4q_parameters}
\begin{tabular}{cccccc}\hline
		Circuits & $P_\text{theo}$ & $d^{(1)}_\text{vigo}$, $d^{(2)}_\text{vigo}$ & $\langle d_\text{vigo}\rangle_\text{theo}$ & $d^{(1)}_\text{athens}$, $d^{(2)}_\text{athens}$ & $\langle d_\text{athens}\rangle_\text{theo}$  \\ \hline
        D4M4$^\dag$ & 0.473 & $86.07$ & 182.07 & 130.87 & 276.85 \\
        D3M4 & 0.390 & $74.00$ & 189.65 & 83.70 & 214.51 \\
        D2M4 & 0.250 & $34.97$ & 139.87 & 47.87 & 191.47 \\
        G1D3M3 & 0.391 & $74.00$ & 189.43 & 83.70 & 214.26 \\
        G2D2M2 & 0.250 & $34.97$ & 139.87 & 46.73 & 186.93 \\
        D4D4M4$^\dag$ & 0.908 & $161.13$ & 177.38 & 250.27 & 275.50 \\
        D3D4M4 & 0.821 & $166.13$ & 202.28 & 207.00 & 252.04 \\
        D2D4M4 & 0.660 & $99.57$ & 150.81 & 140.87 & 213.37 \\
        D4M1$\mid$D3M3 & 0.561 & $72.07$, $74.00$ & 260.35 & 104.80, 83.70 & 336.10 \\
        D3M1$\mid$D3M3 & 0.537 & $74.00$, $74.00$ & 275.53 & 83.70, 83.70 & 311.65 \\
        D2M1$\mid$D3M3 & 0.488 & $34.77$, $74.00$ & 222.75 & 47.80, 83.70 & 269.30 \\
        D4M2$\mid$D2M2 & 0.578 & $72.33$, $34.97$ & 185.61 & 105.20, 46.73 & 262.81 \\
        D3M2$\mid$D2M2 & 0.531 & $74.00$, $34.97$ & 205.09 & 83.70, 46.73 & 245.50 \\
        D2M2$\mid$D2M2 & 0.438 & $34.97$, $34.97$ & 159.85 & 47.93, 46.73 & 216.38\\\hline
	\end{tabular}
\end{table}

\begin{table}[ht]
	\caption{Success probabilities, selectivities and depths of the four-qubit search circuits on IBM's Vigo and Athens machines. Circuit names with $\dag$ are the standard Grover's algorithms. Standard deviations are obtained from $30$ trials with random target states. Each trial has $8192$ shots.}
	\label{Table:4q_results}
\begin{tabular}{ccccccc}\hline
		Circuits & $P_\text{vigo}$ & $S_\text{vigo}$ & $\langle d_\text{vigo}\rangle$ & $P_\text{athens}$ & $S_\text{athens}$ & $\langle d_\text{athens}\rangle$ \\ \hline
        D4M4$^\dag$ & $0.165\pm0.018$ & $1.91\pm0.17$ & $526.66\pm54.17$ & $0.181\pm0.010$ & $2.28\pm0.33$ & $725.65\pm41.46$ \\
        D3M4 & $0.195\pm0.019$ & $2.10\pm0.37$ & $382.49\pm39.23$ & $0.208\pm0.016$ & $1.90\pm0.28$ & $404.27\pm31.08$ \\
        D2M4 & $0.173\pm0.013$ & $2.06\pm0.24$ & $203.71\pm17.62$ & $0.170\pm0.008$ & $1.78\pm0.17$ & $282.61\pm15.28$ \\
        G1D3M3 & $0.209\pm0.018$ & $3.69\pm0.51$ & $356.10\pm44.06$ & $0.230\pm0.010$ & $4.67\pm0.41$ & $364.03\pm11.11$ \\
        G2D2M2 & $0.199\pm0.019$ & $8.43\pm1.99$ & $176.24\pm10.40$ & $0.211\pm0.004$ & $13.36\pm2.76$ & $221.21\pm5.72$ \\
        D4D4M4$^\dag$ & $0.195\pm0.022$ & $2.19\pm0.17$ & $833.99\pm83.83$ & $0.183\pm0.018$ & $2.03\pm0.38$ & $1378.31\pm136.04$ \\
        D3D4M4 & $0.151\pm0.017$ & $1.76\pm0.16$ & $1110.15\pm116.23$ & $0.205\pm0.013$ & $2.44\pm0.40$ & $1013.44\pm64.76$ \\
        D2D4M4 & $0.197\pm0.023$ & $2.15\pm0.20$ & $513.59\pm60.89$ & $0.195\pm0.019$ & $2.41\pm0.31$ & $727.35\pm64.34$ \\
        D4M1$\mid$D3M3 & $0.264\pm0.055$ & $1.46\pm0.16$ & $570.74\pm88.11$ & $0.282\pm0.012$ & $1.58\pm0.06$ & $670.76\pm28.76$ \\
        D3M1$\mid$D3M3 & $0.253\pm0.038$ & $1.46\pm0.18$ & $595.99\pm83.11$ & $0.286\pm0.017$ & $1.64\pm0.11$ & $586.68\pm28.91$ \\
        D2M1$\mid$D3M3 & $0.249\pm0.038$ & $1.40\pm0.13$ & $443.90\pm50.68$ & $0.286\pm0.014$ & $1.64\pm0.10$ & $460.51\pm18.70$ \\
        D4M2$\mid$D2M2 & $0.311\pm0.045$ & $1.58\pm0.17$ & $351.38\pm42.16$ & $0.324\pm0.015$ & $1.87\pm0.17$ & $444.56\pm20.50$ \\
        D3M2$\mid$D2M2 & $0.314\pm0.042$ & $1.68\pm0.23$ & $352.18\pm37.03$ & $0.333\pm0.021$ & $1.78\pm0.19$ & $392.85\pm21.87$ \\
        D2M2$\mid$D2M2 & $0.345\pm0.077$ & $1.70\pm0.19$ & $211.35\pm39.81$ & $0.335\pm0.043$ & $1.70\pm0.37$ & $287.58\pm39.97$ \\\hline
	\end{tabular}
\end{table}

\section{\label{App:B} Supplementary data for average target states}

Calibration parameters of the IBM quantum processors (for the average target states) are listed in Table \ref{Table:machines_average}. The results for the three-, four-, and five-qubit search based on the average target states are listed in Tables \ref{Table:3q_results_average}, \ref{Table:4q_results_average} and \ref{Table:5q_results_average} respectively. 

\begin{table}[ht]
	\caption{Calibration specs for the IBM quantum processors were retrieved on the day of data (for average target states).}
	\label{Table:machines_average}
\begin{tabular}{cccccccc}\hline
		Backends & CNOT error & Readout error & $T_1$ & $T_2$ & Quantum volume & Version & Date  \\ \hline
        Athens & 1.212e-2 & 1.476e-2 & 85.96 & 78.79 & 32 & 1.3.19 & June 9, 2021 \\
        Guadalupe & 1.245e-2 & 2.056e-2 & 78.15 & 90.72 & 32 & 1.2.18 & June 12, 2021 \\\hline
	\end{tabular}
\end{table}

\begin{table}[ht]
	\caption{Success probabilities, selectivities and depths of the three-qubit search with average target states. Circuit names with $\dag$ are the standard Grover's algorithms. Standard deviations are obtained from eight trails (with eight different target states). Each trial has $8192$ shots.}
	\label{Table:3q_results_average}
\begin{tabular}{cccc}\hline
		Circuits & $P_\text{athens}$ ($P_\text{highest}$, $P_\text{lowest}$) & $S_\text{athens}$ & $\langle d_\text{athens}\rangle$  \\ \hline
        D3M3$^\dag$ & $0.526\pm 0.028$ (0.550, 0.470) & $5.38\pm0.78$ & $76.26\pm3.57$ \\
        D2M3 &  $0.370\pm0.018$ (0.380, 0.335) & $2.48\pm0.13$ & $94.93\pm6.56$ \\
        G1D2M2 & $0.441\pm 0.018$ (0.453, 0.425) & $18.02\pm5.60$ & $22.09\pm2.14$ \\
        D3D3M3$^\dag$ & $0.616\pm 0.025$ (0.660, 0.593) & $8.15\pm0.90$ & $110.92\pm2.57$\\ 
        D3M1$\mid$D2M2 & $0.645\pm 0.030$ (0.691, 0.582) & $2.77\pm0.29$ & $92.03\pm3.18$ \\
        D2M1$\mid$D2M2 & $0.611\pm 0.016$ (0.638, 0.594) & $2.26\pm0.14$ & $87.73\pm4.05$ \\\hline
	\end{tabular}
\end{table}

\begin{table}[ht]
	\caption{Success probabilities, selectivities and depths of the four-qubit search with average target states. Circuit names with $\dag$ are the standard Grover's algorithms. Standard deviations are obtained from sixteen trails (with sixteen different target states). Each trial has $8192$ shots.}
	\label{Table:4q_results_average}
\begin{tabular}{cccc}\hline
		Circuits & $P_\text{athens}$ ($P_\text{highest}$, $P_\text{lowest}$) & $S_\text{athens}$ & $\langle d_\text{athens}\rangle$  \\ \hline
        D4M4$^\dag$ & $0.177\pm0.016$ (0.207, 0.157) & $2.15\pm0.46$ & $744.81\pm59.87$ \\
        D3M4 & $0.209\pm0.014$ (0.231, 0.175) & $2.01\pm0.15$ & $401.15\pm25.07$\\
        D2M4 & $0.178\pm0.010$ (0.193, 0.152) &  $1.98\pm0.15$ & $269.56\pm18.53$\\
        G1D3M3 & $0.218\pm0.014$ (0.237, 0.185) & $4.03\pm0.37$ & $384.65\pm22.82$\\
        G2D2M2 & $0.211\pm0.06$ (0.219, 0.200) & $13.60\pm3.62$ & $221.79\pm59.09$\\
        D4D4M4$^\dag$ & $0.163\pm0.014$ (0.196, 0.137) & $1.87\pm0.28$ & $1548.59\pm129.69$\\
        D3D4M4 & $0.219\pm0.015$ (0.253, 0.200) & $2.71\pm0.35$ & $947.37\pm60.44$\\
        D2D4M4 & $0.188\pm0.017$ (0.217, 0.161) & $2.23\pm0.22$ & $757.44\pm65.97$\\
        D4M1$\mid$D3M3 & $0.263\pm0.017$ (0.298, 0.231) & $1.53\pm0.12$ & $720.55\pm45.95$\\
        D3M1$\mid$D3M3 & $0.268\pm0.024$ (0.305, 0.213) & $1.60\pm0.16$ & $628.39\pm56.31$\\
        D2M1$\mid$D3M3 & $0.269\pm0.022$ (0.306, 0.222) & $1.62\pm0.16$ & $490.25\pm36.06$\\
        D4M2$\mid$D2M2 & $0.328\pm0.016$ (0.346, 0.294) & $1.73\pm0.13$ & $463.54\pm25.87$\\
        D3M2$\mid$D2M2 & $0.339\pm0.015$ (0.360, 0.305) & $1.86\pm0.16$ & $385.39\pm15.82$\\
        D2M2$\mid$D2M2 & $0.364\pm0.015$ (0.382, 0.336) & $2.14\pm0.16$ & $259.73\pm11.33$\\\hline
	\end{tabular}
\end{table}

\begin{table}[ht]
	\caption{Success probabilities, selectivities and depths of the five-qubit search with average target states. Circuit names with $\dag$ are the standard Grover's algorithms. Standard deviations are obtained from thirty two trails (with thirty two different target states). Each trial has $8192$ shots.}
	\label{Table:5q_results_average}
\begin{tabular}{cccc}\hline
		Circuits & $P_\text{guada}$ ($P_\text{highest}$, $P_\text{lowest}$) & $S_\text{guada}$ & $\langle d_\text{guada}\rangle$  \\ \hline
		D5M5$^\dag$ & $0.0392\pm0.0606$ (0.0522, 0.0184) & $0.656\pm0.154$ & $3072.73\pm857.30$\\
        G2D3M3 & $0.0570\pm0.0050$ (0.0636, 0.0471) & $1.453\pm0.164$ & $1263.78\pm111.71$\\
        G3D2M2 & $0.0985\pm0.0025$ (0.1038, 0.0942) & $8.971\pm1.355$ & $444.13\pm8.45$\\
        D2M2$\mid$D3M3 & $0.0576\pm0.0109$ (0.0779, 0.0371) & $0.884\pm0.192$ & $2078.96\pm429.64$\\
        D3M3$\mid$D2M2 & $0.1135\pm0.0284$ (0.1568, 0.0647) & $0.804\pm0.310$ & $1093.12\pm322.35$\\\hline
	\end{tabular}
\end{table}


\providecommand{\noopsort}[1]{}\providecommand{\singleletter}[1]{#1}%

\end{document}